\documentclass[onecolumn]{aastex631}

\usepackage{graphicx}
\usepackage[caption=false]{subfig}
\usepackage{booktabs}
\usepackage{amsmath}
\usepackage[thinc]{esdiff}

\usepackage{rotating}

\defcitealias{Wyper2024}{Paper~I}

\hypersetup{linkcolor=magenta,citecolor=blue,filecolor=cyan,urlcolor=purple}

\newcommand\editone[1]{\textcolor{black}{#1}}

\shorttitle{Synthetic Observations of a Pseudostreamer CME}
\shortauthors{Lynch et al.}

\begin{document}

\title{Synthetic Remote-sensing and In-situ Observations of Fine-scale Structure\\ in a Pseudostreamer Coronal Mass Ejection through the Solar Corona}

\correspondingauthor{Benjamin~J.~Lynch}
\email{bjlynch@ucla.edu}

\author[0000-0001-6886-855X]{B.~J.~Lynch} 
\affiliation{Department of Earth, Planetary, and Space Sciences, University of California--Los Angeles, Los Angeles, CA 90056, USA}
\affiliation{Space Sciences Laboratory, University of California--Berkeley, Berkeley, CA 94720, USA}

\author[0000-0002-6442-7818]{P.~F.~Wyper} 
\affiliation{Department of Mathematical Sciences, Durham University, Durham, DH1 3LE, UK}

\author[0000-0001-6590-3479]{E.~Palmerio} 
\affiliation{Predictive Science Inc., San Diego, CA 92121, USA}



\author[0000-0002-5068-4637]{L.~Casillas}
\affiliation{Department of Earth, Planetary, and Space Sciences, University of California--Los Angeles, Los Angeles, CA 90056, USA}

\author[0000-0002-9493-4730]{J.~T.~Dahlin} 
\affil{Astronomy Department, University of Maryland, College Park, MD 20742, USA}
\affil{Heliophysics Science Division, NASA Goddard Space Flight Center, Greenbelt, MD 20771, USA}

\author[0000-0002-1198-5138]{L.~K.~S.~Daldorff} 
\affil{The Catholic University of America, N.E. Washington, DC 20064, USA}
\affil{Heliophysics Science Division, NASA Goddard Space Flight Center, Greenbelt, MD 20771, USA}


\author[0000-0003-1439-4218]{S.~E.~Guidoni} 
\affiliation{Department of Physics, American University, Washington, DC 20016, USA}
\affil{Heliophysics Science Division, NASA Goddard Space Flight Center, Greenbelt, MD 20771, USA}

\author[0000-0003-1380-8722]{A.~K.~Higginson} 
\affil{Heliophysics Science Division, NASA Goddard Space Flight Center, Greenbelt, MD 20771, USA}


\author[0000-0001-6289-7341]{P.~Kumar} 
\affiliation{Department of Physics, American University, Washington, DC 20016, USA}
\affil{Heliophysics Science Division, NASA Goddard Space Flight Center, Greenbelt, MD 20771, USA}

\author[0000-0002-0016-7594]{A.~Liberatore}
\affil{Jet Propulsion Laboratory, California Institute of Technology, Pasadena, CA 91109, USA}

\author[0000-0002-5068-4637]{P.~C.~Liewer}
\affil{Jet Propulsion Laboratory, California Institute of Technology, Pasadena, CA 91109, USA}

\author[0000-0002-4440-7166]{O.~Panasenco}
\affil{Advanced Heliophysics, Pasadena, CA 91106, USA}

\author[0000-0001-6759-2037]{P.~Penteado}
\affil{Jet Propulsion Laboratory, California Institute of Technology, Pasadena, CA 91109, USA}

\author[0000-0002-2381-3106]{M.~Velli}
\affil{Department of Earth, Planetary, and Space Sciences, University of California--Los Angeles, Los Angeles, CA 90056, USA}




\begin{abstract}

Coronal pseudostreamer flux systems have a specific magnetic configuration that influences the morphology and evolution of coronal mass ejections (CMEs) from these regions. 
Here we continue the analysis of the \citeauthor{Wyper2024}\ (\citeyear{Wyper2024}, ApJ \textbf{975}, 168) magnetohydrodynamic simulation of a CME eruption from an idealized pseudostreamer configuration through the construction of synthetic remote-sensing and in-situ observational signatures. 
We examine the pre-eruption and eruption signatures in extreme ultraviolet and white-light from the low corona through the extended solar atmosphere. 
We calculate synthetic observations corresponding to several Parker Solar Probe-like trajectories at $\sim$10$R_\odot$ to highlight the fine-scale structure of the CME eruption in synthetic WISPR imagery and the differences between the in-situ plasma and field signatures of flank and central CME-encounter trajectories.  
Finally, we conclude with a discussion of several aspects of our simulation results in the context of interpretation and analysis of current and future Parker Solar Probe data. 
 
\end{abstract}



\section{Introduction} \label{sec:intro}

The launch and ongoing operation of the Parker Solar Probe \citep[PSP;][]{Fox2016} mission, combined with the increase in solar activity, has resulted in a wealth of opportunities for coordinated multispacecraft observations of coronal mass ejection (CME) eruptions and their evolution through the extended corona and inner heliosphere \citep{Velli2020}. One of the primary scientific goals of PSP is to characterize the corona--heliospheric connection, which is especially relevant for CMEs and other coherent, transient heliospheric structures such as streamer blobs, reconnection jet outflows, magnetic switchbacks, etc \citep{Raouafi2023}. The PSP instrumentation is primarily for in-situ observations---with its own magnetic field \citep[FIELDS;][]{Bale2016}, plasma \citep[SWEAP;][]{Kasper2016}, and energetic particle measurements \citep[IS$\odot$IS;][]{McComas2016}---however, it also includes white-light (WL) heliospheric imagery from the two Wide-field Imager for Solar Probe Plus \citep[WISPR;][]{Vourlidas2016} telescopes.

The overall, large-scale WL structure of CMEs (the three-part, bright front, dark cavity, and bright central/trailing core morphology) and its correspondence to a large-scale magnetic flux rope-like geometry is relatively well-understood \citep{Illing1985,Dere1999,Cremades2004,Vourlidas2013,HowardTA2017,Lynch2016,Lynch2021}. 
The PSP/WISPR heliospheric imaging of CMEs and other large transients now routinely captures the dynamics and evolution of fine-scale WL structures and substructures throughout different regions of the ejecta.
Some of the first analyses of slow, streamer-blowout type of CMEs in PSP/WISPR data were presented by \citet{HowardR2019,Hess2020}, and shortly thereafter a variety of 3D tracking techniques for CMEs were developed and applied \citep{Braga2021,Liewer2021}. There has been an increasing number of impressive observations and analyses of fine-scale structure associated with CMEs and their adjacent flows. For example, \citet{HowardR2022} investigated the evolution and dissolution of the CME leading edge into multiple fronts and a fair amount of overlapping internal substructure in the CME ``core'' region. \citet{Shaik2024} have examined the nested-ring WL structures of CMEs, showing they were consistent with concentric flux surfaces. The distortion of the WL CME cavity shape (flux rope cross-section) during its interaction with the structured background wind was investigated by \citet{Braga2022}. And recently, \citet{Cappello2024} have examined Large Angle Spectroscopic Coronagraph \citep[LASCO;][]{Brueckner1995} observations taken by the Solar and Heliospheric Observatory \citep[SOHO;][]{Domingo1995} in conjunction with simultaneous PSP/WISPR imaging of a CME event that exhibits both complex internal structure and an entire train of highly structured, intermittent post-CME outflows. We note, these examples complement and expand an entirely different set of WISPR observations of smaller-scale, slow wind streamer blob and heliospheric current sheet/plasma sheet transient outflow structures \citep[e.g.][]{Rouillard2020,Reville2022,Poirier2020,Poirier2023,Ascione2024,Liewer2024}.

\citet{Wyper2024}, hereafter \citetalias{Wyper2024}, performed a detailed magnetohydrodynamic (MHD) simulation with the Adaptively Refined MHD Solver \citep[ARMS;][]{Devore2008} of the energization and eruption of a pseudostreamer CME with a particular emphasis on magnetic reconnection and its impact on the CME's dynamic structure and connectivity during different phases of the eruption. They identified four primary reconnection scenarios over the course of the eruption: ($i$.)~bursty, breakout reconnection outflows removing overlying restraining flux prior to the CME onset; ($ii$.)~the ``classic'' 3D eruptive flare and flux rope formation reconnection; ($iii$.)~the subsequent disconnection of one leg of the CME flux rope through interchange reconnection; and ($iv$.)~the post-eruption relaxation and rebuilding of the pseudostreamer flux system. 
Given the \citetalias{Wyper2024} analysis and interpretation of the variety of roles magnetic reconnection can play throughout the eruption's development, our aim with the current study is to relate the simulation's magnetic field dynamics with potential observational signatures.
In a completely analogous sense, the present work is to \citetalias{Wyper2024}, as \citet{Lynch2004} was to \citet{MacNeice2004}: a complementary analysis via synthetic extreme ultraviolet (EUV) and/or WL observations derived from the MHD simulation data from different viewpoints.

In fact, the organization of the paper largely follows that of \citet{Lynch2004}. 
In the first part of the study, we characterize the CME eruption in the low corona with synthetic EUV imaging (section~\ref{sec:overview}) and throughout the extended corona in synthetic WL data (section~\ref{sec:middle}), and suggest the coronal morphology and height--time kinematics can be related to measures of magnetic reconnection during the eruption sequence. 
Additionally, in Section~\ref{sec:interp}, we present a visualization of the 3D magnetic field evolution and its correspondence to the structure and dynamics in the synthetic EUV and WL imagery during each of the four phases of magnetic reconnection used to characterize the pseudostreamer CME eruption in \citetalias{Wyper2024}, including: $\S$\ref{sec:interp:one}, the ``standard flare'' flux rope formation and eruption; $\S$\ref{sec:interp:two}, the CME leg disconnection; and $\S$\ref{sec:interp:three}, the pre- and post-eruption periods of interchange reconnection between the pseudostreamer flux system and the surrounding open field. 
In the second part of the study (Section~\ref{sec:insitu}), we present the first simultaneous, forward-modeled synthetic PSP/WISPR WL heliospheric imaging alongside the time-dependent in-situ bulk plasma and magnetic field sampling \citep[adapted from][]{Lynch2022} for three different PSP orbit--impact trajectories with the CME's magnetic ejecta: a remote-sensing non-encounter; a flank encounter; and a central encounter. And lastly, in Section~\ref{sec:insitu:wispr}, we present a qualitative morphological comparison and simulation-inspired interpretation of PSP/WISPR observations of helical substructure in the trailing outflow of the 2020 Jan 26 CME event. 
We conclude, in Section~\ref{sec:disc}, with a discussion of our results and the prospects for future modeling and their detailed comparisons with observations.

\section{Eruption Overview---EUV Low Corona} \label{sec:overview}

The eruption onset and low-coronal evolution of the pseudostreamer CME is described in detail in \citetalias{Wyper2024}. These simulation results largely followed the general development and eruption phases seen in our prior modeling \citep[e.g.,][]{Lynch2013,Wyper2017,Masson2019,Wyper2021}. 
There are a number of observational papers that have looked at pseudostreamer CME eruptions in EUV or a combination of EUV and WL. 
For example, the low-coronal EUV structure and morphology during the early phases of the eruption process were analyzed by \citet{Mason2021} and \citet{Kumar2021} discussed the low-coronal EUV signatures for a number of pseudostreamer CMEs. We will refer to specific figures in each of those works in the following sections describing our synthetic EUV imaging results.

The detailed procedure for calculating our line-of-sight integrated synthetic EUV intensities from different viewpoints is outlined in Appendix~\ref{sec:method} along with an explanation of the imaging processing pipeline used herein. Due to our simplified treatment of the MHD energy equation, our synthetic EUV emission is approximated as $I_{\rm EUV} \sim \int d\ell \, n_e^2$ allowing us to avoid having to calculate a model EUV spectrum from the appropriate emission lines (at each point along the line of sight) or convolve this integrated simulated emission with specific instrument filter response functions \citep{Downs2010}. We conjecture that synthetic EUV emission structures would be \emph{even more complex} than what we present here if the plasma's self-consistent thermodynamic response to field-aligned heat flux, radiative losses, and coronal heating terms were included in the MHD energy equation.

\subsection{Transition from Quasi-stable Cavity to CME Eruption}

\begin{figure*}
    \centering
    \includegraphics[width=0.96\textwidth]{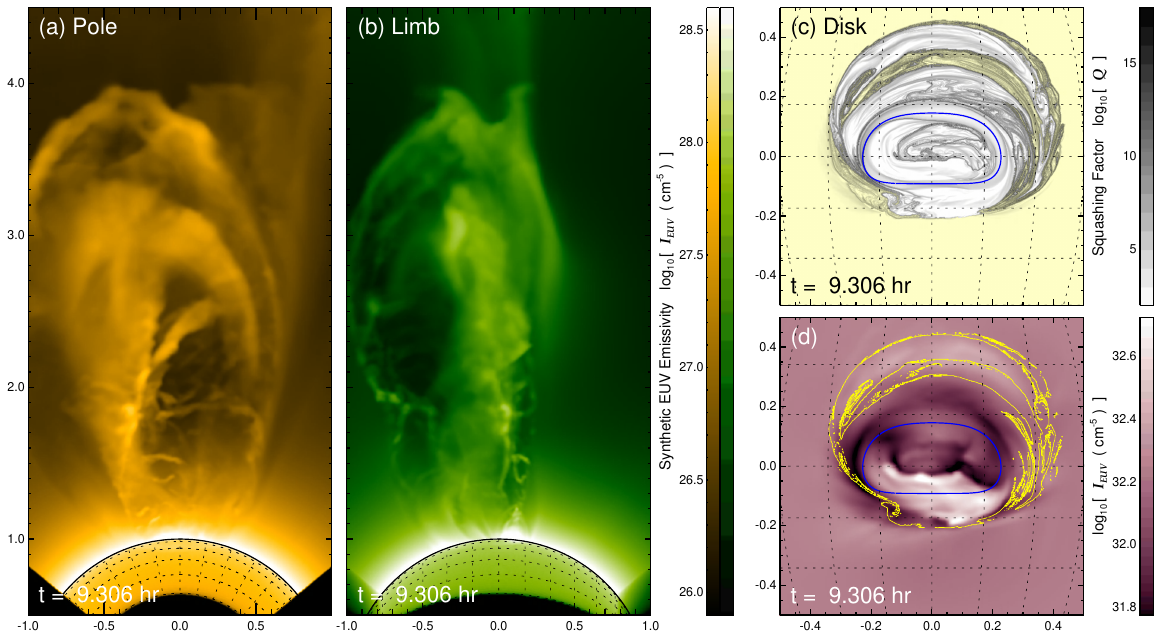}
    \caption{Synthetic EUV emission and magnetic field complexity of a pseudostreamer CME eruption from different observational configurations. Panels (a) \emph{Pole} and (b) \emph{Limb} show off-limb EUV observations from their respective viewpoints, while panel (d) \emph{Disk} shows the central, on-disk EUV morphology. Panel (c) shows the $B_R=0$ contour (dark blue) over $\log{Q}$, a measure of magnetic field geometric complexity, with the open flux regions shaded light yellow. The open--closed boundary is shown as the yellow line in panel (d). The (a) and (b) color scales are identical and used here to distinguish the two viewpoints.\\(An animation of this figure is available.)}
    \label{fig:euv}
\end{figure*}

Figure~\ref{fig:euv} shows synthetic EUV emission signatures during the early stages of the eruption capturing the explosive large-scale re-structuring of the magnetic field and the rapid formation of fine-scale density structure in response. 
The different viewing geometries shown in the figure are \ref{fig:euv}(a) \emph{Pole}: looking down at the pseudostreamer CME from the solar north pole, \ref{fig:euv}(b) \emph{Limb}: view from the ecliptic plane where the center of the pseudostreamer lies directly in the plane of the sky.), and \ref{fig:euv}(c),(d) \emph{Disk}: view from the ecliptic plane at the central $\phi=0^{\circ}$ meridian (for more detail, see Appendix~\ref{sec:method}).
The synthetic EUV dynamics leading up to the eruption onset are complicated due to the simultaneous (or near-simultaneous) development of multiple physical processes in the pre-eruption to eruption phase transition \citep{Patsourakos2020}. 
In our configuration, both breakout reconnection (at the stressed separatrix boundary and pseudostreamer null) and the classic tether-cutting/eruptive flare reconnection (below the erupting flux rope) processes are contributing to the pseudostreamer eruption dynamics and CME evolution \citep{Karpen2012,Lynch2016a}.

\citet{Kumar2021} have discussed the structure and evolution of breakout reconnection in pseudostreamer eruptions in both numerical simulations and recent EUV observations.
In the animation of Figure~\ref{fig:euv} for times $t \lesssim 8.5$~hr, we see this breakout reconnection process accelerating with increased upflows and downflows coming out of the extended, overlying current layers formed at the separatrix boundary. The magnetic island plasmoids generated in the current sheet have a complex 3D structure, as seen by comparing the corresponding bright features at the same height from the two viewing perspectives. From the edge-on \emph{Limb} view, the plasmoid flux ropes appear as compact, blob-like intensity concentrations, whereas from the face-on \emph{Pole} view of the reconnecting current layers, these enhancements appear as sheet-like arcs or extended collimated features. 
We expect this is completely analogous to the 3D structure of larger, less-energetic streamer blob magnetic island plasmoids observed and modeled from helmet streamers \citep[e.g.,][and references therein]{Higginson2018,Lynch2020,Reville2022},
and we know that extended, coherent island structures within the current layer largely reflect the relative magnitude of the guide field component \citep{Edmondson2017}.

The $t \lesssim 8.5$~hr period also corresponds to the ``slow rise phase" \citep{ZhangJ2004} of the pre-eruption sheared core/twisted flux rope. 
The development of the CME essentially follows the evolution described in Section~4.1 of \citet{Masson2019}. While the formation of the hyperbolic flux tube occurs relatively early in the simulation ($t\sim6$~hr), the continued arcade expansion and sheared/twisted core flux rising leads to the formation of the classic eruptive flare current sheet \citep{Janvier2013} and establishes the positive feedback between arcade/flux rope expansion and increased reconnection at the growing eruptive flare current sheet.
Note that in animation of Figure~\ref{fig:euv}, in the frames from $t=8.889$~hr up to and including $t=9.306$~hr, the 
\ref{fig:euv}(a) \emph{Pole} EUV morphology is reminiscent of the SDO/AIA 131~{\AA} observations in Figure~9 of \citet{Kumar2021}, including the overlying leading-edge loop front and both roll and twisting motion in the CME material above the activated side arcade during eruption \citep{Panasenco2011}. Likewise during this period, the \ref{fig:euv}(b) \emph{Limb} eruption morphology shows similarities with the SDO/AIA 193~{\AA} observations in Figure~7 of \citet{Kumar2021}, specifically in the asymmetry of the erupting half of the pseudostreamer, and the overall CME deflection towards the null point \citep{Lynch2013,Sahade2022}. 
The synthetic EUV structure of the material that will become the ``core'' in the traditional 3-part CME \citep{Vourlidas2013} is partially scooped up with the rising flux rope but also introduced into the erupting flux rope structure by the flare reconnection processes. It is precisely due to this reconnection-generated fine-scale structure that our synthetic emission shows such qualitative agreement with high-resolution EUV observations of prominence eruptions, e.g., compare Figure~\ref{fig:euv}(a) with Figure~3(d) in \citet{Mason2021}.

\subsection{On-disk Signature of Dynamic Topological Evolution}

Figure~\ref{fig:euv}(c) shows the evolution of the logarithmic squashing factor, $\log{Q}$ \citep{Titov2007}, which is a geometric measure of structure and complexity in the magnetic field's connectivity mapping, and sometimes referred to as the $Q$-map.  
A complete description and annotation of the $Q$-map evolution during each phase of the eruption can be found in Section~4.2 and Figure~12 of \citetalias{Wyper2024}. 
For our purposes here, the general features of interest are the open and closed flux designations---open flux regions are shaded yellow---and the general morphology and complexity of the $\log{Q}$ values shown in grayscale.
In the animated version of Figure~\ref{fig:euv}, the $Q$-map evolution for $t < 8.75$~hr is largely closed-flux separatrix boundary moving slowly \citep{Lynch2008,Lynch2013} but beginning to develop whirl/swirl substructure on this boundary as the reconnection becomes increasingly plasmoid-dominated \citep{Wyper2021b,Dahlin2022a}. The additional structure that appears over the 500~s interval between the $t=8.889$ and 9.028~hr frames shows the inherent complexity of the pseudostreamer's closed flux opening up, i.e.\ there is a dynamic mixing of the connectivity \citep[e.g.,][]{Higginson2017a} throughout the entire eruption until finally the pseudostreamer flux system is essentially restored for $t \gtrsim 11.8$~hr.

Figure~\ref{fig:euv}(d) shows the evolution of synthetic EUV emission looking down on the pseudostreamer source region from the \emph{Disk} viewpoint (Appendix~\ref{sec:method}). The polarity inversion line (where the radial component of the field $B_R$ is zero) is shown as dark blue, and the instantaneous open--closed boundary derived from the panel (c) $Q$-map is shown as the yellow contours. Given that our synthetic EUV emission proxy has a dependence on $n_e^2$, here the darker (brighter) regions reflect real line-of-sight integrated density depletion (accumulation). Despite the relative simplicity of our EUV emission approximation, the density evolution that is reflected in our synthetic EUV imagery and Figure~\ref{fig:euv} animation captures: 1) the development of CME-related transient dimming regions \citep{Dissauer2018,Dissauer2019,Krista2022,Ngampoopun2023}; 2) the fainter outlying dimming due to the material outflow from interchange reconnection \citep[discussed further in Section~\ref{sec:interp:three}; see also][]{Wyper2022}; and 3) the formation and growth of the eruptive flare arcade and its circular ribbon front \citep{Karpen2024, Palmerio2025}.
Flare emission can be used as an indirect proxy of reconnection flux \citep[e.g.,][]{Kazachenko2017,Gopalswamy2017}. Therefore, it is a useful exercise to forward-model the synthetic evolution of flare arcades and ribbons to verify this relationship with the numerical simulation results \citep{Lynch2019,Lynch2021,Dahlin2022a,Kazachenko2022}. 
One advantage of the rigorous topology analysis with flux system boundaries determined via $\log{Q}$ values is that it yields the most accurate quantification of the magnetic flux and its rate of change. Below we will refer to the \citetalias{Wyper2024} definitions of the following quantities: $\Phi_{\rm FR}$ as the CME flux rope azimuthal/twist flux component; $\Phi_{\rm INT}$ as the total flux having undergone interchange reconnection; $d\Phi_{\rm RXN}/dt$ as the closed-field reconnection rate calculated from the area swept by the flare ribbons; and $d\Phi_{\rm INT}/dt$ as the interchange reconnection rate representing open--closed reconnection.

\section{Eruption Overview---White Light Extended Corona} \label{sec:middle}

\subsection{Morphology and Evolution of the CME Structure and Substructure}

CME morphology in the corona, \editone{historically, has generally been} categorized as ``clearly a flux rope,'' corresponding to the well-known three-part CME structure observed in coronagraph imaging data \citep{Bothmer1998,Dere1999,Sheeley1999,ChenJ2000,Cremades2004} or ``clearly not a flux rope,'' which covers almost everything else, i.e.\ narrow jet-like CMEs, diffuse wave fronts, or intermittent, bursty transient outflows \citep[e.g.,][]{Vourlidas2013, Vourlidas2017,Nitta2021}. \editone{By the STEREO era (2006--present), the increase in white-light imaging resolution as well as simultaneous multi-point viewing led to the understanding that this categorization dichotomy was largely artificial; almost all CMEs exhibit some form of complexity and their white-light morphology can be significantly impacted by the observing geometry and line-of-sight integration effects \citep{Vourlidas2017}.}
Since coronagraph observations show a tremendous range of morphological features with various levels of complexity, an especially important feature of MHD simulations is the ability to analyze the true 3D structure of the magnetic field and entrained plasma density structures. 
This appears to be even more the case for CMEs originating from coronal pseudostreamer source regions, where the structure and evolution of eruptive transients are significantly influenced by the surrounding magnetic environment \citep{Panasenco2011,Kumar2021,Sahade2022,Karpen2024}.
For example, \citet{Wang2015} and \citet{WangYM2018} have discussed the role of pseudostreamer topology in forming a variety of non-traditional white light eruptive signatures, including narrow, jet-like and broader, fan-shaped CMEs, and \citet{Kumar2021} have illustrated a variety of observational signatures ranging from breakout reconnection dimming and outflows to the CME's rolling and un-twisting motions during the eruption. Recently, \citet{WangYM2023} analyzed several pseudostreamer CME eruptions that showed evidence of lateral confinement by the adjacent open fields and signatures of interchange reconnection opening up the erupting magnetic flux rope structure.

Previous modeling efforts, including deriving synthetic observable signatures in EUV and WL, have reported that moderately complex magnetic field configurations can correspond to relatively simple/smooth density variations \citep{Lynch2016a,Lynch2021} while relatively simple magnetic field configurations can generate highly complex field and plasma structures and substructures \citep[e.g.,][]{Wyper2016b,Dahlin2022a}.  
Here we have used the Thomson-scattering formalism \citep{Billings1966} to calculate time series of synthetic white light intensity images constructed from the integration of arrays of line-of-sight traced through the 3D simulation density distribution. See Appendix~\ref{sec:method} for further details and the image processing procedure.

\begin{figure*}
    \centering
    \includegraphics[width=0.96\textwidth]{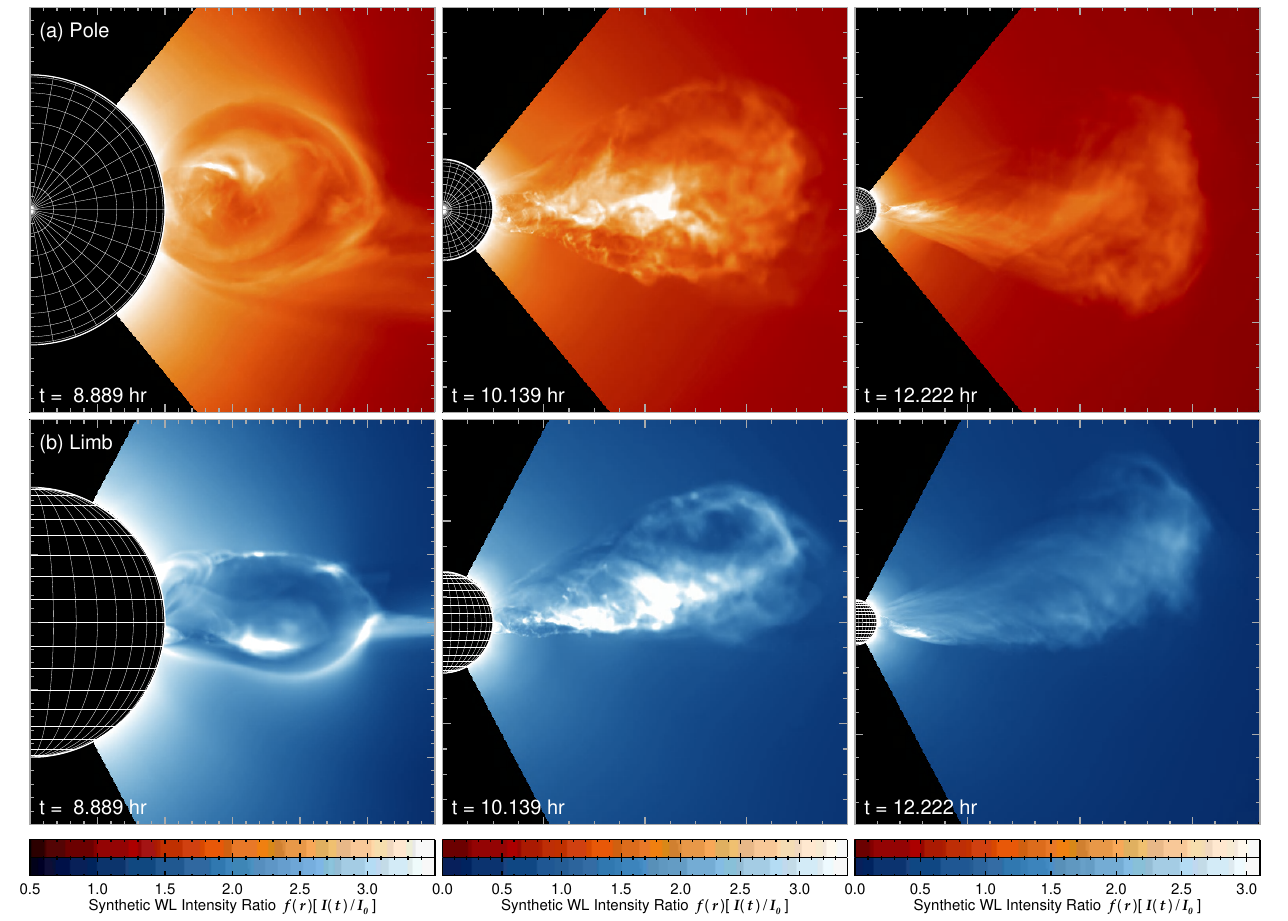}
    \caption{Evolution of the pseudostreamer CME and its morphology in synthetic Thomson-scattered white light observations from the (a) \emph{Pole} and (b) \emph{Limb} viewpoint configurations. The synthetic WL images are normalized by the $I_0$ profile of a spherically symmetric, gravitationally-stratified density fall off (see Appendix \ref{imgproc}). The two color scales are identical and are used here to distinguish the two viewpoints. \\(An animation of this figure is available.)}
    \label{fig:wl0}
\end{figure*}

Figure~\ref{fig:wl0} shows synthetic Thomson-scattered WL emission derived from the \citetalias{Wyper2024} MHD data cubes for two viewpoints, \ref{fig:wl0}(a) \emph{Pole} and \ref{fig:wl0}(b) \emph{Limb}, at three times during the eruption. One of the most immediately noticeable features of Figure~\ref{fig:wl0} is the amount of internal substructure in the reconnection exhaust occurring in the wake of the main CME eruption. The synthetic coronagraph observations in the middle column ($t=10.14$~hr) show how unstructured the ``core'' of the CME becomes by $\sim$5$R_\odot$---in contrast to the classic, well-structured three-part CME at $t=8.89$~hr. 
\editone{The animation of Figure~\ref{fig:wl0} shows the CME complexity becoming seemingly less complex and less bright (with respect to the background intensity) for $t \gtrsim 10$~hr as it propagates through the domain. This is due to a combination of the location and intensity of the post-CME material outflow changing in time, the existing density structures experiencing 3D spherical expansion and potentially additional CME over-expansion, and therefore, the density structures' contributions to the line-of-sight scattering integrals are spread further from the plane-of-the-sky or Thomson-sphere point of maximum scattering efficiency.}
Nevertheless, the extended, less-energetic phase of the eruptive flare reconnection that is acting to rebuild the closed flux system in the wake of the main eruption continues to impart structured, bursty outflows into the solar wind behind the impulsive eruption of the twisted/sheared flux. We also note the eruption is somewhat more narrow when viewed in the \emph{Limb} perspective ($\sim$30$^\circ$ width) compared to the broader \emph{Pole} viewpoint ($\sim$45$^\circ$ width)---reflecting the slightly larger longitudinal extent of the active region flux distribution. 

\subsection{Eruption Kinematics}

\begin{figure*}[!t]
    \centering
    \includegraphics[width=0.96\textwidth]{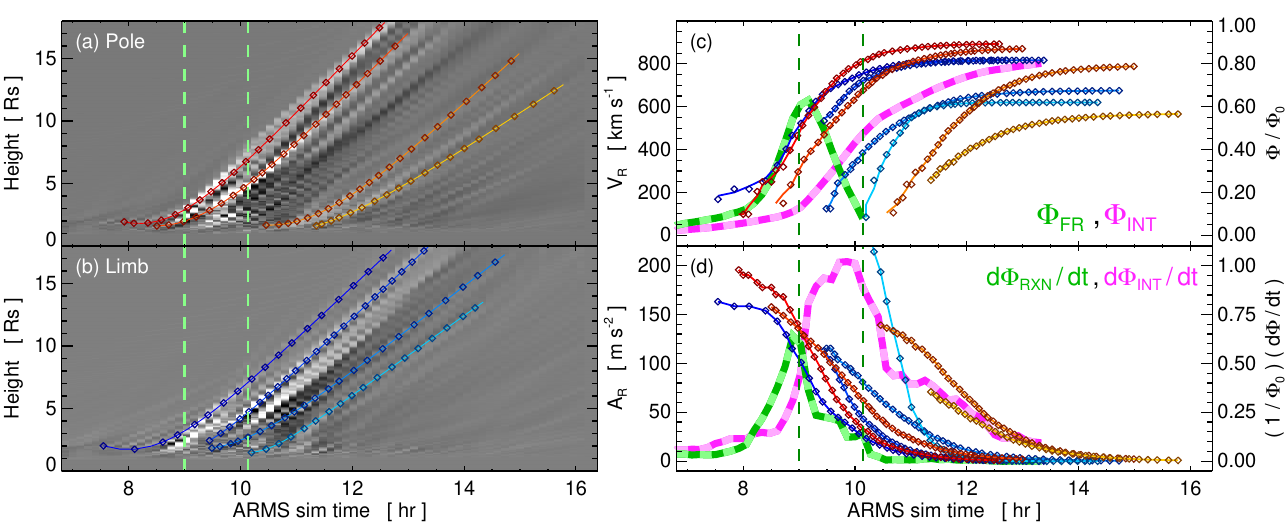}
        \caption{Kinematic profiles derived from the running-difference height-time plots sampled at a position angle of $90^{\circ}$ for the (a) \emph{Pole} and (b) \emph{Limb} image sequences shown in the animation of Figure~\ref{fig:wl0}. Panels \editone{(c)} and \editone{(d)} show the calculated $V_{R}(t)$ and $A_{R}(t)$ profiles, respectively, for each of the height-time tracks in (a), (b) in the same colors. The normalized CME flux rope content $\Phi_{\rm FR}$ and the total interchange reconnection flux $\Phi_{\rm INT}$ are shown in (c) while the associated (normalized) flare-ribbon flux and interchange reconnection rates ($d\Phi_{\rm RXN}/dt$, $d\Phi_{\rm INT}/dt$) are shown in (d). The vertical dashed green lines show the period between the onset of fast eruptive flare reconnection through the CME flux rope leg disconnection.}
    \label{fig:wl1}
\end{figure*}

Since the discovery of slow and fast CME velocity profiles in coronagraph images \citep{Sheeley1999}, there has been considerable interest in relating the height, velocity, and acceleration profiles of CMEs with the initiation mechanism or other physical processes---such as magnetic reconnection---associated with the eruption \citep[e.g.,][]{ZhangJ2001,ZhangJ2004,Gallagher2003,Jing2005,Qiu2004,Qiu2010,Welsch2018,Zhu2020,Vievering2023}. 
Figure~\ref{fig:wl1}(a) and \ref{fig:wl1}(b) show the height--time plots generated from running-difference processing of the the $18 R_\odot \times 18 R_\odot$ WL image sequence from  Figure~\ref{fig:wl0} for the \emph{Pole} and \emph{Limb} viewpoints, respectively. Representative fits to the height--time points are shown as the red-to-orange and blue-to-cyan colored lines.
The height--time curves and the running difference tracks represent the \emph{apparent plane-of-the-sky position} of features in the synthetic WL image sequence rather than, e.g.\ the 3D radial distance of points on a specific field line or of plasma features on a specific plane, as were used in \citetalias{Wyper2024}. The \emph{Pole} view plane-of-the-sky corresponds to the $\theta=90^\circ$ equatorial plane while the \emph{Limb} plane-of-sky corresponds to the $\phi = 0^\circ$ meridian. The Figure~\ref{fig:wl1} running difference slices are constructed from an average of 1D samples in the $x$-direction centered around the Sun's center ($y$, $z = 0$).
We follow the \citet{Lynch2021} procedure to fit the height--time profile data with the \citet{Sheeley1999} function
\begin{equation}
    h(t) = r_0 + 2 r_a \ln{ \left[ \cosh{ \left[ \frac{v_a \left( \; t+t_0 \; \right)}{2 r_a} \right] } \right] } \; ,
    \label{eq:ht}
\end{equation}
which yields the following velocity and acceleration profiles as a function of radial distance (height),
\begin{equation}
    V_R( \; h(t) \; ) = v_a \left( 1 - \exp{ \left[ -\frac{( \; h(t)-r_0 \;)}{r_a} \right] } \right)^{1/2} \; , \;\;\;
    A_R( \; h(t) \; ) = \left( \frac{v_a^2}{2 r_a} \right) \exp{ \left[ -\frac{( \; h(t)-r_0 \;)}{r_a} \right] } \;\; .
\end{equation}
The best fit is obtained via the IDL {\tt curvefit} routine to minimize the weighted-$\chi^2$ between Equation~\ref{eq:ht} model values and the height-time points obtained via the usual point-and-click IDL user-interface.  
The set of best-fit parameters, $\left[ \, t_0, \, r_0, \, r_a, \, v_a \right]$, for each of the \emph{Pole} and \emph{Limb} height--time tracks are listed in Table~\ref{table:s1}. The most relevant parameter is the asymptotic velocity, $v_a$, which can be considered the final speed of each curve fit. 
Figure~\ref{fig:wl1}(c) plots the $V_R(t)$ profiles for each of the height--time tracks in the color scheme used in panels (a),(b), and Figure~\ref{fig:wl1}(d) shows the corresponding $A_R(t)$ profiles.

%
\begin{table*}[t!]
\begin{tabular}{|ll|l|cccc|c|}
     \hline
     &  & type & $t_0$ {[}hr{]} & $r_0$ {[}$R_\odot${]} & $r_a$ {[}$R_\odot${]} & $v_a$ {[}km s$^{-1}${]} & Std. Error $\sigma_{\rm fit}$ \\
     \hline
     %
     %
\underline{\textbf{Pole Viewpoint}} 	& track 1 & CME & -8.193   &   1.805   &   2.971  &    895.11 & 0.1143\\
Figure~\ref{fig:wl1}a     			& track 2 & CME & -8.485   &   1.613   &   3.581  &    877.06 & 0.1017\\
(red to orange)     				& track 3 & post-CME &  -10.511   &   1.669   &   3.330    &  795.23 & 0.0725 \\
     							& track 4 &  post-CME &  -10.490   &   1.027   &   2.597   &   568.88 & 0.0406 \\
     \hline
     %
     %
\underline{\textbf{Limb Viewpoint}} 	& track 1 & CME & -7.986   &   1.720   &   2.816    &  819.44 & 0.1176\\
Figure~\ref{fig:wl1}b    			& track 2 & CME & -8.569   &   1.148   &   2.418    &  817.41 & 0.0614\\
(blue to cyan)     				& track 3 & post-CME &  -9.154   &   1.706   &   2.736   &   676.58 & 0.1059\\
     							& track 4 &  post-CME &  -10.098    &  1.456    &  1.063   &   620.25 & 0.0531 \\
     \hline
\end{tabular}
\caption{Best-fit parameters for the Equation~\ref{eq:ht} $h(t)$ profiles shown in Figure~\ref{fig:wl1}. \label{table:s1}}
\end{table*}

The running-difference height--time tracks show the same general plane-of-outflow evolution over the course of the eruptive flare as seen in the animations of Figures~\ref{fig:euv} and \ref{fig:wl0}. Here, the brightest tracks first appear in the \emph{Pole} view and then at the \emph{Limb} before returning to the \emph{Pole} view as fainter, slower outflows well after the eruption. This temporal ordering of the height--time tracks is the exact ordering of the best-fit asymptotic velocity parameter---the first two red curves and the first two blue curves (the `CME' track type) have final speeds ranging from 820--900~km~s$^{-1}$ whereas the second set of orange and cyan curves (the `post-CME' tracks) show a broader deceleration trend with final velocities between 550--800~km~s$^{-1}$ (also seen in the clustering of the $V_R(t)$ profiles in panel \ref{fig:wl1}(c)). 
We also note these extended, material outflows that continue long after the CME eruption, e.g.\ through $t \sim 13$~hr in the \emph{Pole} height--time tracks, are precisely the WL counterpart to the post-eruption interchange reconnection jet outflows seen in the synthetic EUV imagery and discussed further in Section~\ref{sec:interp:three}.

While there is general statistical agreement that greater reconnection flux generally implies a more energetic (and therefore, faster) CME \citep{Qiu2004,Qiu2005}, each individual CME eruption will likely have its own unique velocity and acceleration profiles, reflecting in part, the influence of source region's magnetic configuration and the adjacent and overlying streamer flux systems and the specific details of the energy release processes \citep{Zhu2020,Vievering2023}. If the flare reconnection is a significant driver of the CME's early evolution, then one may expect the well-known Neupert  effect \citep{Neupert1968,Qiu2021} relating $I_{\rm SXR} \sim \int dt \, I_{\rm HXR}$ to have some connection to the flare reconnection $\Phi_{\rm RXN} \sim \int dt \left( d\Phi_{\rm RXN} / dt \right)$ and to the CME's $V_R \sim  \int dt \, A_R$ profile. 
\editone{In fact, semi-analytical models of Lorentz force-driven flux rope propagation that have explicitly related CME velocity to flux content, i.e.~$A_R(t) \propto d\Phi_{\rm RXN}/dt$, have been shown to produce kinematic profiles generally consistent with the observed CME's height-time evolution \citep{ChenJ2000,ChenJ2010}.}

Figure~\ref{fig:wl1} includes several measures of magnetic reconnection occurring during the eruption process. The vertical line at $t=9$~hr denotes the onset of fast eruptive flare reconnection and the flux rope eruption and the line at $t=10.28$~hr marks the full  disconnection/opening-up of one CME leg. 
In panel \ref{fig:wl1}(c) we plot the closed-flux content of the CME flux rope, $\Phi_{\rm FR}/\Phi_0$, as the dashed green curve and the flux having undergone interchange reconnection, $\Phi_{\rm INT}/\Phi_0$, as the dashed magenta curve. The normalization is set to the total closed-field flux $\Phi_0 = 6.243 \times 10^{21}$~Mx. Here, the $V_R(t)$ profiles of the four `CME' tracks rise in tandem with the reconnection flux into the CME flux rope $\Phi_{\rm FR}$ and roll over towards near constant values of $\gtrsim 800$~km~s$^{-1}$ while the four `post-CME' tracks show decreasing asymptotic velocities.
In panel \ref{fig:wl1}(d), we show the normalized eruptive flare reconnection rate calculated from the rate of change of the flare reconnection flux (the magnetic flux swept by flare ribbon area, $\Phi_0^{-1} \left( d\Phi_{\rm RXN}/dt \right)$ as the dashed green curve and the normalized interchange reconnection rate, $\Phi_0^{-1} \left( d\Phi_{\rm INT}/dt \right)$, as the dashed magenta curve.
The $A_R(t)$ profiles for the four `CME' tracks smoothly decrease at roughly the same time as flare-ribbon flux reconnection rate ($d\Phi_{\rm RXN}/dt$), whereas the four `post-CME' $A_R(t)$ profiles are spread out further in time and the latter three more closely follow the decay of the interchange reconnection rate ($d\Phi_{\rm INT}/dt$) curve. The separation appears to correspond to the $t=10.28$~hr line indicating the full disconnection of one CME leg, representing the completion of the expulsion (opening up) of the energized low-lying flux into the open field.


\section{Reconnection-generated Fine-scale Structure in Synthetic Observations}
\label{sec:interp}

In \citetalias{Wyper2024}, we identified four primary reconnection phases over the course of the pseudostreamer CME eruption: the eruptive flare reconnection and CME flux rope formation, the interchange reconnection disconnection of one of the CME legs, and the pre-eruption and post-eruption periods of interchange reconnection. 
In this section, we extend the analysis of \citetalias{Wyper2024} through illustrative examples of reconnection-generated fine-scale structure in the synthetic EUV and/or WL observations during each of these reconnection phases and discuss their relationship to the dynamic magnetic field reconfiguration and evolution.

\subsection{Flux Rope Formation and White-light Cavity Substructures} \label{sec:interp:one}

\begin{figure*}
    \centering
    \includegraphics[width=0.96\textwidth]{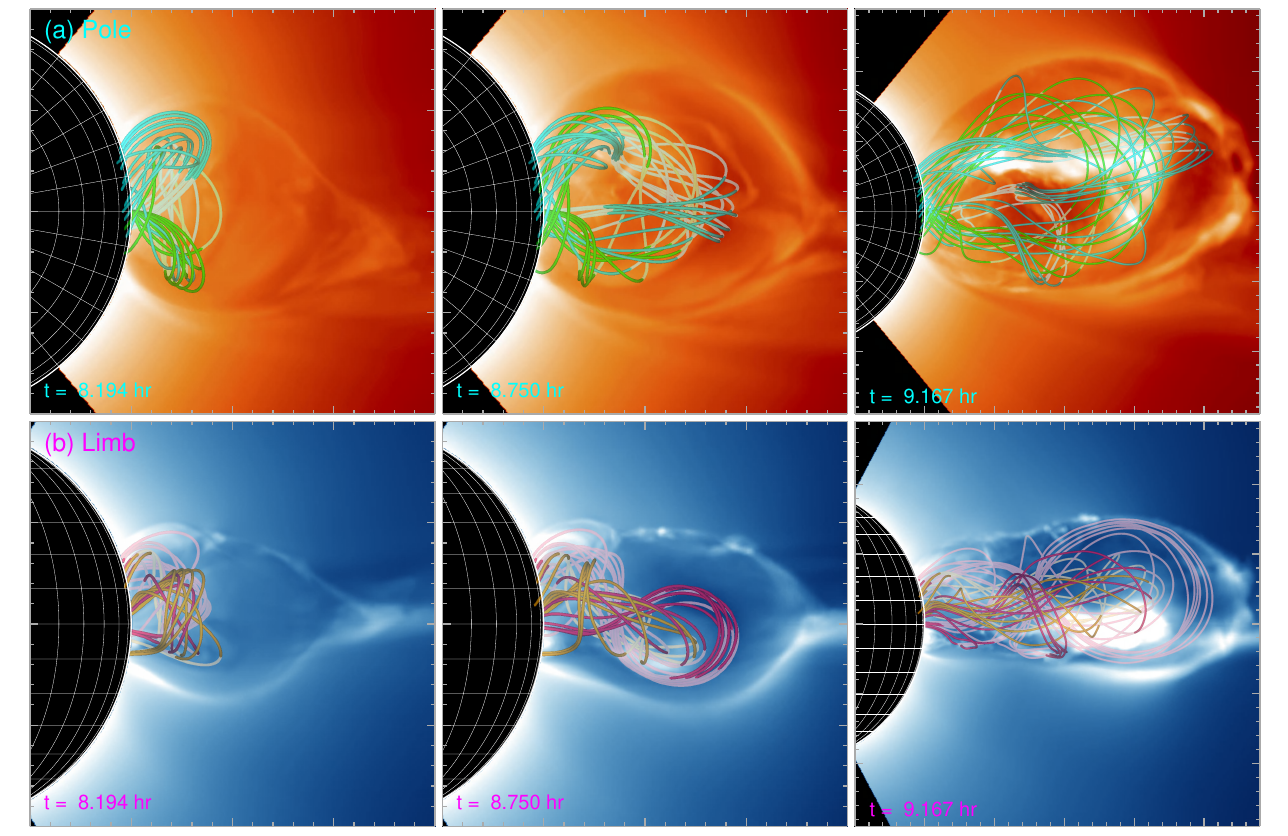}
        \caption{Visualization of the structure and evolution the eruption of the CME's energized core flux rope fields from the (a) \emph{Pole} and (b) \emph{Limb} viewpoints using the synthetic Thomson-scattered white light intensity images. The same representative field lines are plotted in each panel, traced from the line-tied inner boundary, thus accurately depicting the field line's temporal evolution. The green and cyan field lines in row (a) are shown as the orange and purple field lines, respectively, in row (b).}
    \label{fig:FRfls}
\end{figure*}

Understanding the structure and evolution of coronal cavities may be a way to estimate energized, pre-eruptive magnetic field configurations and their stability criteria \citep[e.g.,][]{Gibson2010,Guennou2016,Karna2019,Kumar2021}.
There continues to be some ambiguity in the association of specific plasma features with estimates or interpretations of the magnetic field dynamics and how these signatures evolve from the lower to the outer corona \citep{Vourlidas2013}. In Figure~\ref{fig:FRfls}, we have combined the 3D magnetic field line visualizations with the synthetic Thomson-scattered WL imagery during the formation and initial eruption of the pseudostreamer CME flux rope from the (a) \emph{Pole} and (b) \emph{Limb} viewpoints. The set of field lines at each time are traced from the same set of positions on the line-tied lower radial boundary to show the true field and connectivity evolution. Likewise, the cyan (green) field lines in the top row are the same field lines in magenta (orange) shown in the bottom row.

When we compare the synthetic WL structure of the $t=8.194$~h \emph{Limb} panel with the magnetic topology visualizations presented Figure~13 of \citetalias{Wyper2024} ($\log Q$, average twist $T_w$), we see a faint local dimming outlining the flux rope, the brighter pseudostreamer separatrix boundaries, and the transient dynamics of interchange reconnection opening up the restraining overlying pseudostreamer flux. The bright cusp at the reconnection outflow along the pseudostreamer spine in the \emph{Limb} panel is seen to have a more extended triangular shaped enhancement from the \emph{Pole} perspective.  
As the eruption develops, the mass accumulation seen in the core by $t=8.75$~h is being supplied by the onset of fast eruptive flare reconnection \citep{Lynch2016a}. By $t=9.17$~h, the bright core structure appears as a peanut-shaped blob from the \emph{Limb} perspective, extending from the center to the trailing region of the CME flux rope cross-section. However, from the \emph{Pole} viewpoint, we see a well-resolved reconnection jet structure with collimated (field-aligned) density enhancements along successive, adjacent flux bundles. This is simply the (complex) 3D version of the flare reconnection-supplied density enhancement often used to define the flux rope core in axisymmetric CME simulations \citep[e.g.,][]{Lynch2004,Riley2008,Zuccarello2012}. 
Lastly, we note that the by $t=9.17$~h panel, the dark cavity portion of the synthetic WL observations is now completely encompassed by the CME flux rope, whereas in the middle panel the flux rope was a smaller proportion of the cavity region which still included some overlying pseudostreamer flux that had not yet reconnected out of the way.

\subsection{CME Flux Rope Leg Disconnection} \label{sec:interp:two}

CME interaction with the surrounding and adjacent flux systems during eruption has implications for homologous and sympathetic eruptions \citep{Devore2008,Lynch2013}, solar energetic particle acceleration and transport \citep{Masson2013,Masson2019}, the motion of adjacent arcades and the flux rope connectivity \citep{Dudik2019,Zemanova2019}, and even governing of flux circulation between the ejecta its surroundings \citep{Titov2022}.
This CME--adjacent field reconnection early in a CME's low-coronal evolution was seen in simulation results and discussed by \citet{Lugaz2011}, and, as detailed in \citetalias{Wyper2024}, there are series of comprehensive analyses of pseudostreamer CMEs' morphology and dynamics in the low corona as signatures of ongoing reconnection processes \citep{WangYM2018,WangYM2023,Kumar2021}.
In general, CME interaction with the ambient heliosphere has been of considerable interest. For example, it was determined that interchange reconnection acting on the erupting closed-fields of CMEs was necessary to avoid the so-called heliospheric flux budget pile-up catastrophe \citep{Crooker2002, Crooker2012, Reinard2004}. And from the in-situ magnetic field and plasma observations at 1~au and elsewhere, there is ample evidence for continual reconnection during a CME's heliospheric propagation that results in the erosion of the coherent, closed-field connectivity of the CME structure \citep{Dasso2006,Ruffenach2015,Pal2022}.

\begin{figure*}[t]
    \centering
    \includegraphics[width=0.96\textwidth]{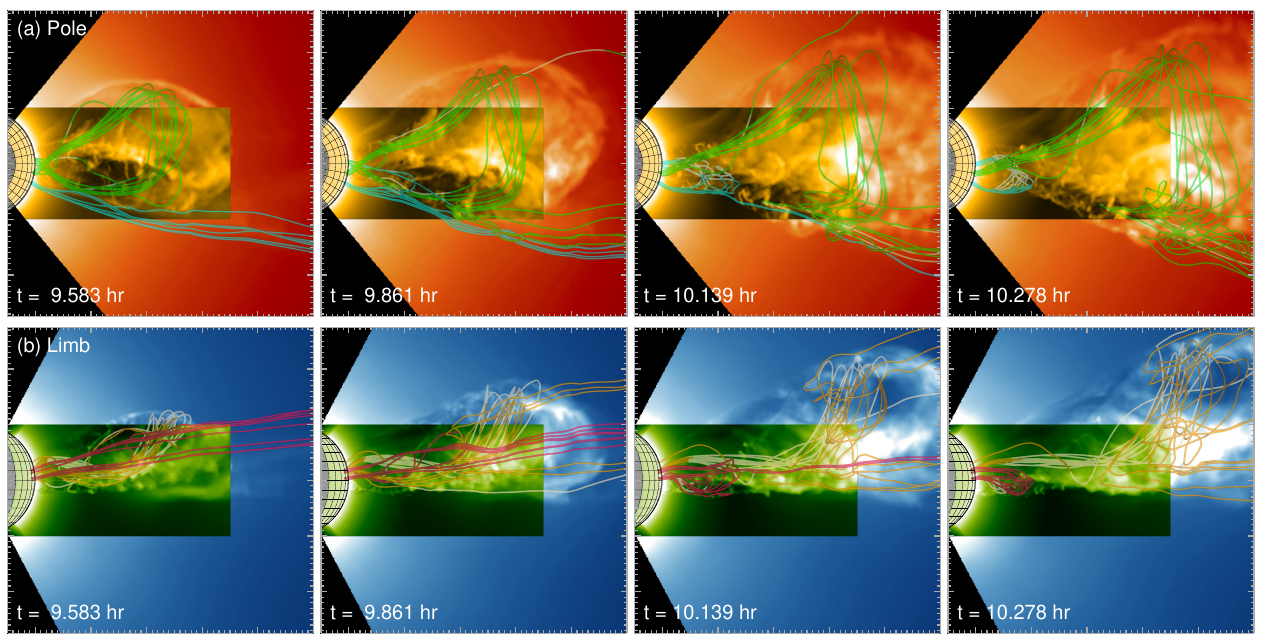}
        \caption{Selected composite frames of synthetic EUV (central rectangle) and white light (background) from the row (a) \emph{Pole} and row (b) \emph{Limb} viewpoints, showing the magnetic reconnection process between one CME leg and the adjacent open fields with representative field lines. The same field lines are shown in each of the rows with the green--orange and cyan--red color correspondence used in Figure~\ref{fig:FRfls}. Here, the green and orange CME field lines were previously closed and open up while the cyan and red field lines were previously open and close down to become part of the pseudostreamer flux system.}
    \label{fig:rxnfls}
\end{figure*}

Once the CME flux rope has formed and is erupting, the magnetic topology of its pseudostreamer source region continues to play a critical role in the eruption's development. We refer the reader to the full description of the CME leg disconnection provided in Section~3.3 of \citetalias{Wyper2024} and illustrated in their Figure~5. 
To summarize, during the eruption, the overlying breakout current sheet on the surface of the separatrix dome and the eruptive flare current sheet beneath the CME flux rope eventually combine into a single exterior current layer surrounding the flux rope, which allows the null point (reconnection site) to move within the sheet. As the CME flux rope expands, one leg will have its magnetic field orientation in the same direction as the surrounding unipolar open field while the other CME leg will have the opposite polarity, resulting in anti-parallel fields on either side of the combined, exterior flux rope current layer. The flux rope leg disconnection occurs via interchange reconnection at this current sheet.

In Figure~\ref{fig:rxnfls}, we show a composite image of the synthetic EUV data from Figure~\ref{fig:euv} over the corresponding synthetic WL data from Figure~\ref{fig:wl0}. We also include representative magnetic field lines calculated via the same methodology as in Figure~\ref{fig:FRfls}, with the the green (cyan) field lines in the \emph{Pole} viewpoint panels corresponding to the yellow (red) field lines in the \emph{Limb} panels. Our aim with the magnetic field line sequences was to reproduce the topological evolution shown in Figure~5 of \citetalias{Wyper2024} and we note the selection of simulation output times here includes the first and last frames ($t=9.583$ and 10.139~h) of the \citetalias{Wyper2024} figure.
The initial internal density substructure developing in the core region in the Figure~\ref{fig:FRfls}(a) panels is seen here as the continued expansion the reconnection jet outflow that continues to supply mass density to the newly-formed, outermost layers of the erupting CME flux rope. The upper (westward) leading edge of the erupting structure in the synthetic WL \emph{Pole} view is seen as a natural extension of its corresponding feature in the synthetic EUV emission. However, the lower (eastward) CME flux rope boundary is much more dynamic as the location of the interchange reconnection that erodes the CME's connectivity and eventually opens up the entire eastern leg of the CME flux rope ejecta. The exchange of open--closed connectivity in Figure~\ref{fig:rxnfls} is shown as the previously-closed green CME field lines reconnecting with the adjacent open flux cyan field lines leading to the closing down of the cyan flux as the green CME flux opens up, generating dynamic fine-scale outflow structures---seen in both modeling and observations \citep{Wyper2016b,Kumar2019}.

\subsection{Pre- and Post-CME Interchange Reconnection Outflows} \label{sec:interp:three}

\begin{figure*}[t]
    \centering    \includegraphics[height=0.98\textwidth,angle=90]{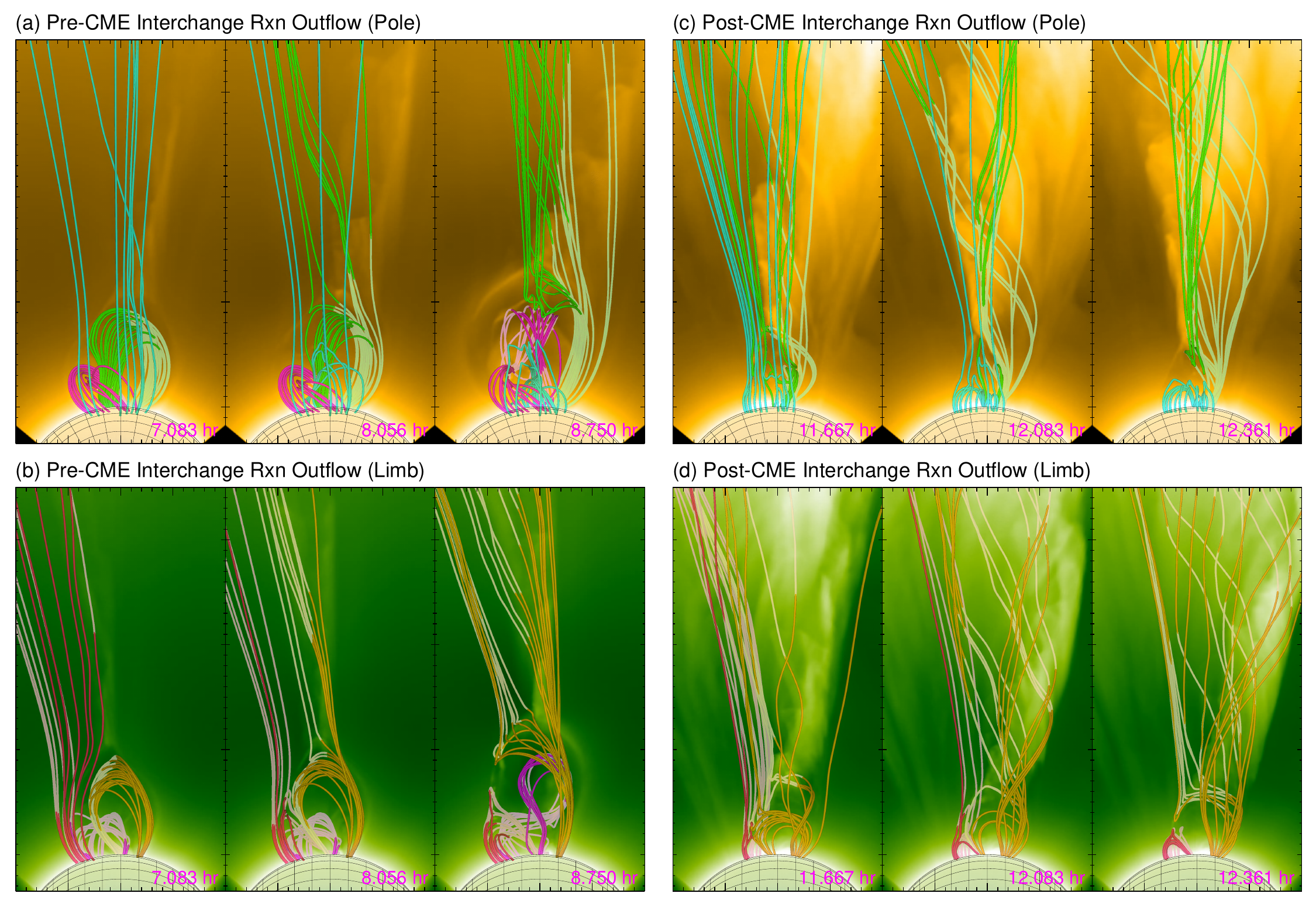}
        \caption{Select frames showing the two distinct, extended phases of interchange reconnection. {Left column}, the pre-eruption, magnetic breakout interchange reconnection from the (a) \emph{Pole} and (b) \emph{Limb} viewpoints; {Right column}, the post-eruption interchange reconnection that rebuilds the pseudostreamer flux system from the (c) \emph{Pole} and (d) \emph{Limb} viewpoints. The same representative field lines are plotted in each of the viewpoint pairs in the style of Figure~\ref{fig:rxnfls}.}
    \label{fig:IRsigs}
\end{figure*}

Interchange reconnection has long been associated with a number of dynamic coronal phenomena \citep[e.g.][and references therein]{Lynch2014, Masson2014, Higginson2017b, Edmondson2017, Scott2021, Scott2022, Wyper2022, Pellegrin-Frachon2023}. 
The two examples of interchange reconnection discussed here, despite occurring before and after the main eruptive transient, happen in the same overall geometric configuration representing the MHD relaxation process of releasing the stored magnetic stress energy. 
\citetalias{Wyper2024} described the pre-eruption interchange reconnection in Section~3.1 and the post-eruption interchange reconnection in Section~3.4. 
Figure~\ref{fig:IRsigs} shows the first period of pre-eruption breakout reconnection that occurs largely before the CME eruption from the \ref{fig:IRsigs}(a) \emph{Pole} and \ref{fig:IRsigs}(b) \emph{Limb} viewpoints in the synthetic EUV imagery along with the representative magnetic field lines illustrating the topological evolution. Here the opening up of the pseudostreamer closed flux system is shown as the evolution of the green (orange) field lines alongside the simultaneous closing down of the open flux shown as the evolution of the cyan (red) field lines. 
Both the field and plasma signatures of these pre-eruption breakout reconnection outflows are seen in the EUV signatures and show dynamic transient outflows along the outer spine and fan structure (\citealt{Lynch2013, Masson2012, Kumar2019, Kumar2021, Wyper2017, Wyper2022}; \citetalias{Wyper2024}). The collimated density enhancement outflow from the \emph{Limb} perspective is seen to be more of an extended sheet with propagating ripples from the \emph{Pole} viewpoint. \citet{Wyper2022} performed an ultra-high resolution, adaptively-refined simulation of the \citetalias{Wyper2024} plasmoid-unstable pre-eruption interchange reconnection and its outflow into the solar wind and discussed density enhancements spreading out along successively-reconnected flux bundles in addition to the intermittent, bursty release of torsional Alfv\'{e}n waves associated with successive magnetic island plasmoid merging into the open flux.

In panels \ref{fig:IRsigs}(c), (d), we show the final interchange reconnection phase in which the remaining, still-conneced CME leg ``jumps'' over the pseudostreamer flux system. We have selected the three simulation times, $t=11.667$, 12.083, and 12.361~h, and traced the field lines inspired by panels (d)--(f) of Figure~7 in \citetalias{Wyper2024}, which essentially corresponds to our \emph{Limb} view, just looking from the opposite direction (toward $-\mathbf{\hat{y}}$ rather than toward $+\mathbf{\hat{y}}$). 
Again, the bright EUV jet signature is seen to correspond to density enhancement spreading out along successively-reconnected field lines as in \citet{Wyper2022}. 
One immediately noticeable difference in Figure~\ref{fig:IRsigs} is the material enhancement with the post-CME interchange reconnection jet outflows is significantly stronger (higher intensity) and covers a greater spatial extent than the analogous outflows in the pre-eruption case. This may be expected given the external, system-driven nature of CMEs, jets, and other explosive eruption scenarios \citep{Antiochos1999,Karpen2012,Wyper2017}, and there is ample observational and simulation support for the generation of intermittent, complex outflows in post-CME trailing current sheets \citep{Riley2007,Karpen2012,Webb2016,Cappello2024}.

CME flux rope field lines exhibit a helical shape and may be expected to show ``unwinding'' motion as CME twist flux propagates/spreads out over the expanding structure. 
In both \ref{fig:IRsigs}(c) and (d), the curvature in the EUV density fronts reflect the large-scale curvature of the magnetic field. These fronts are seen to propagate in tandem as part of the reconnection jet's bulk outflow away from the pseudostreamer current layers into the open field and solar wind.
Additionally, the magnetic reconnection outflow geometry introduces a sharp bend in every newly-reconnected field line which is then seen to propagate away as the field line straightens out. The spatial scale (amplitude) of the field perturbation with respect to the $\boldsymbol{\hat{r}}$ direction is on the order of the size of the separatrix dome $L_x$ which, during periods of sustained interchange reconnection, would be associated with a reconnection-generated \emph{source} of low-frequency Alfv\'{e}nic fluctuations \citep{Lynch2014} of a fundamentally larger spatial/longer temporal scale than, e.g.\ the coherent fluctuations introduced by the intermittent ejection of the much smaller-scale magnetic island plasmoid flux ropes formed in the separatrix current layers, impacting the open field \citep[as in][]{Wyper2022}. 
The Q-map signatures of post-reconnection flux bundles can be seen in panel (c) of the animation of in Figure~\ref{fig:euv}. In particular, the final frame ($t=12.5$~hr) shows the surface area of open flux adjacent to the reformed, closed-flux pseudostreamer system that has undergone interchange reconnection and retain sharp, highly-structured gradients in their connectivity. 
Being able to relate the spatial scale(s) of the pseudostreamer flux system to the strength, location, or other properties of the associated reconnection-generated Alfv\'{e}nic and other MHD waves may be a promising avenue for solar wind acceleration models \citep{Cranmer2018}, source region identification and/or heliospheric back-mapping \citep[e.g.][and references therein]{Lynch2023}.
%

\section{Synthetic Parker Solar Probe Observations}
\label{sec:insitu}

In this section of the study, we present a set of simultaneous remote-sensing and in-situ synthetic observations from a time-dependent observer within the MHD simulation domain. Here we apply the method of \citet{Lynch2022} to derive three orbital trajectories along different intervals of the predicted PSP ephemeris for Encounter 23\editone{---as a representative orbit with the closest perihelion of 9.86$R_\odot$---}and generate the in-situ bulk plasma and magnetic field time series for each. We expand upon our previous analysis by generating the corresponding time series of synthetic white-light heliospheric images as would be seen by the PSP/WISPR Inner and Outer cameras at the location of our virtual spacecraft. 
Our three trajectories increase their proximity to the CME structure: the first trajectory is a remote-sensing only configuration; the second trajectory represents a flank encounter where the spacecraft impacts the outer portion of the magnetic ejecta and its surrounding CME material; the third trajectory is a central encounter where the original PSP latitudes have been offset to ensure a spacecraft intersection with the center of the cavity/strong-field portion of the CME.
The addition of the synthetic WL WISPR imaging from the vantage point of real PSP orbits to our analysis largely follows the procedure outlined in \citet{Poirier2020,Poirier2023} who have examined modeling results of the slow solar wind and helmet streamer structure, including for transient streamer blobs. The details of the synthetic Parker observation calculations are given in Appendix~\ref{sec:psp}. To the best of our knowledge, we are reporting the first demonstration of this capability for a 3D MHD simulation of a self-consistent, physics-based CME eruption throughout the extended solar corona.

The first several well-observed PSP CME events were of the slow, streamer blowout type \citep{Vourlidas2018} that showed various features consistent with a magnetic flux rope-like morphology in the remote-sensing observations \citep{HowardR2019, Hess2020, Wood2020, Liewer2021} as well as flux rope-like field and plasma signatures in the CME in-situ measurements \citep{Korreck2020, Lario2020, Nieves-Chinchilla2020, Palmerio2021c}. One of the main objectives of our earlier study \citep{Lynch2022} was to examine the synthetic in-situ sampling of large-scale, relatively idealized flux rope CME encounters and slowly introduce more ambiguity with the relative spacecraft trajectory orientation with respect to the large-scale CME flux rope axis.
As we have seen above, our pseudostreamer CME simulation generates considerable fine-scale structure in the synthetic coronal imaging observations and, as we will show below, much of this density structure survives into the WISPR heliospheric imaging fields of view. The \editone{in-situ field and plasma structures} are significantly more complex than those examined in \citet{Lynch2022}, which gives us the opportunity to make precisely the opposite point than our previous \editone{conclusions: sometimes analytic flux rope model fits for the observed in-situ field rotations during CME-related transients may not be an accurate representation of the real 3D field structure.} 
For example, a limitation of the current practice analytic flux rope fitting is that it is extremely difficult to distinguish between large-scale magnetic field \emph{twist} and large-scale magnetic field \emph{writhe} from a single observing trajectory through an ordered, coherent magnetic field rotation \citep{AlHaddad2011, AlHaddad2019}. And it has been shown that flank encounters of CME flux rope structures are particularly challenging to constrain via these fitting techniques \citep{Lynch2022,Palmerio2024}.

\subsection{Trajectory 1: Remote-sensing Coverage of CME}
\label{sec:insitu:traj1}

The first PSP trajectory is an interval of the Encounter~23 orbit that allows the CME to propagate through the full WISPR field of view. We line up the ARMS simulation time $t=0$~hr with the PSP ephemeris date and time of 2025-03-22 02:45:00~UT. Additionally, we apply a constant longitudinal offset between the PSP coordinates (given here in Heliocentric Carrington Inertial coordinates) with the ARMS domain's spherical coordinate $\phi_{\rm ARMS}(t) = \phi_{\rm PSP}(t) - 40^\circ$ in order to line up the computational spherical-wedge with the region of the Sun the PSP orbit transverses.

\begin{figure*}[t]
    \centering
    \includegraphics[width=0.96\textwidth]{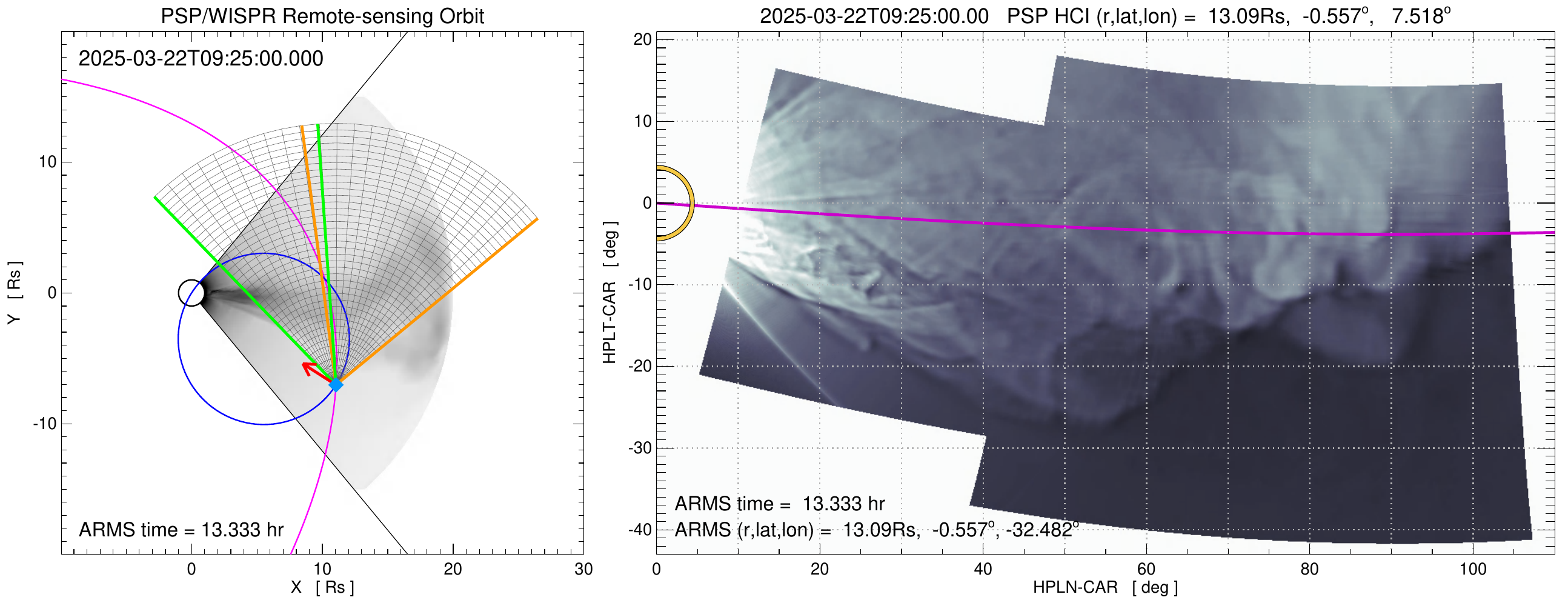}
    \includegraphics[width=0.96\textwidth]{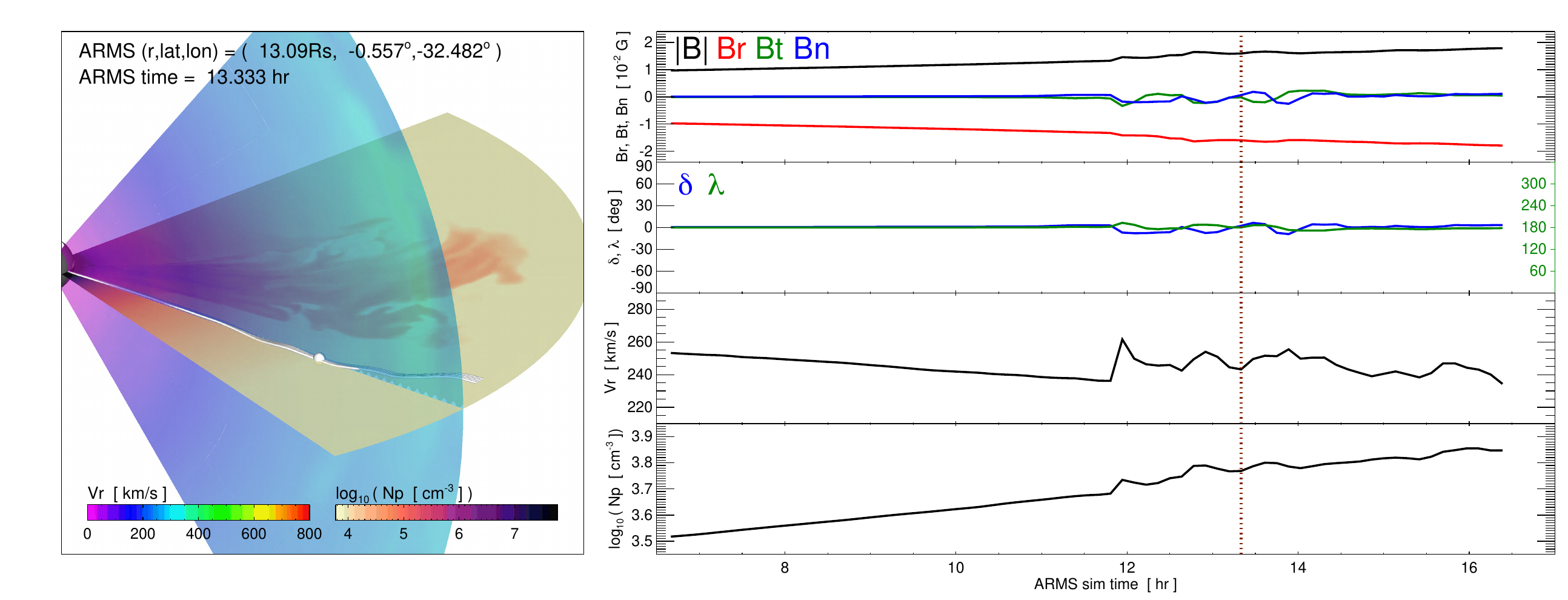}
    \caption{Synthetic PSP observations for the Trajectory~1 remote-sensing encounter with the pseudostreamer CME. Upper left: spacecraft and CME positions in the ecliptic plane from the global \emph{Pole} viewpoint. Upper right: Synthetic white-light heliospheric imaging based on the PSP/WISPR fields of view and camera pointing. Lower left: 3D visualization of MHD solution showing the observing position, radial velocity $V_R$ and number density $N_p$ in meridional and latitudinal planes, respectively. Lower right: synthetic in-situ measurements of the magnetic field and bulk plasma properties from the observing spacecraft. \\(An animation of this figure is available.)}
    \label{fig:psp1}
\end{figure*}

The top row of Figure~\ref{fig:psp1} illustrates the synthetic WISPR imaging procedure. The left panel shows the simulation domain and synthetic WL (inverse grayscale) from the \emph{Pole} viewpoint looking down on the ecliptic, along with the instantaneous position of the PSP-observer spacecraft (blue diamond), the encounter trajectory (magenta), and the angular extent of the WISPR-Inner (green) and WISPR-Outer (orange) heliospheric imagers. The $-\mathbf{\hat{r}}$ direction from PSP is shown as the red arrow and the dark blue circle intersecting the Sun center and the PSP position represents the Thomson sphere. The UTC time for the Parker Encounter 23 orbital position is shown in the upper left while the corresponding simulation time is denoted in the lower left of the plot window. 
The upper right panel of Figure~\ref{fig:psp1} shows the synthetic Thomson-scattered WL intensity in helioprojective longitude and latitude angular coordinates $(\theta_x, \theta_y)$ using the PSP orbit and WISPR camera pointing information. 
The Sun is shown centered at $(0,0)$ with a radius the size of the angle subtended from the PSP distance. The magenta curve shows the PSP orbital plane intersection with the PSP-centered helioprojective plane of the sky. In the animated version of Figure~\ref{fig:psp1}, the temporal position of the observing spacecraft is shown as the eruption proceeds and the WISPR field of view and its observing geometry are updated accordingly.
Appendix~\ref{sec:psp:two} describes the WISPR coordinate deprojection procedure and the simple image processing pipeline applied to our synthetic observables.  

While the CME front enters the WISPR-Inner field of view at $t=9.167$~hr, the collimated, pre-eruption interchange reconnection outflow \citep{Lynch2013,Kumar2021,Wyper2021} is visible at an angle above the ecliptic $\theta_y=0^{\circ}$ plane, a seemingly straight forward extension of the corresponding structures in the synthetic EUV and coronagraph \emph{Limb} perspective imagery (e.g.\ Figures~\ref{fig:FRfls}(b), \ref{fig:IRsigs}(b)). The CME shows a number of concentric (nested) rings in the cavity-to-leading edge front region, whereas the core enhancement is more complex through $t \sim 10.3$~hr. Afterwards, the entire CME ejecta becomes increasingly unstructured, appearing more like an extended, jet-like transient outflow than a flux rope CME.  
This may be the CME version of the outer corona--inner heliosphere ``flocculation,'' \citep{DeForest2016}, however, recall that most of the fine-scale synthetic intensity structures examined in Section~\ref{sec:interp} have been associated with relatively smooth and well-ordered magnetic fields, even shortly thereafter the CME leg disconnection at $t=10.278$~hr in Figure~\ref{fig:rxnfls}.

Figure~\ref{fig:psp1}, lower left, shows the 3D visualization of the MHD simulation data. Specifically, we show the planar cuts of $N_p$ and $V_r$ scalar quantities on semi-transparent latitudinal and longitudinal planes, respectively. The inner radial boundary displays $B_R$ in grayscale as well as the PIL location as the $B_R=0$ contour.
The virtual PSP spacecraft position is shown with the small white sphere. The instantaneous magnetic connectivity of our synthetic observer is illustrated via a flux bundle of five magnetic field lines drawn from spacecraft position and at $\pm1$ and $\pm2$ $\Delta \phi = 0.025^{\circ}$ steps in longitude.  
\editone{We note that far from the CME transient, the field lines are essentially radial and therefore somewhat obscured by the latitudinal and longitudinal planes' intersection in this particular visualization.}
Figure~\ref{fig:psp1}, lower right, shows synthetic in-situ time series generated from the in-situ sampling along the adjusted PSP orbit positions at each of the simulation output times. From top to bottom we plot the magnetic field magnitude $|\mathbf{B}|$ (black) and its Radial--Tangential--Normal (RTN) components $B_R$ (red), $B_T$ (green), and $B_N$ (blue), followed by the magnetic field's local latitudinal and azimuthal pointing angles ($\delta$ blue; $\lambda$ green), the radial velocity $V_R$, and finally the proton number density $N_p$. The vertical dashed line shows the instantaneous time in the orbit interval and is seen moving from left to right in the animated version of Figure~\ref{fig:psp1}.
At $t=11.944$~hr the observing spacecraft grazes a very weak shock, $\Delta V \sim 20$~km~s$^{-1}$.
\editone{This shock structure/discontinuity ahead of the CME is seen in the WISPR images. The WL transient reaches $\theta_x \sim 90^{\circ}$ at precisely the same time as the in-situ measurements show a slight increase in density. The $\theta_x \sim 90^{\circ}$ point represents a local maximum---at the position of the observer---of the WL Thomson-scattering efficiency, so some correspondence is to be expected. Here, the weak shock structure is also seen as the spherical front in the $V_R$ meridional plane of the 3D visualization.} 
After impact with the disturbance ($t \gtrsim 12$~hr), the in-situ time series includes small-scale fluctuations in the plasma and field quantities, seemingly consistent with sampling the boundary layers of a disturbed wake.

\subsection{Trajectory 2: Flank Encounter with CME}
\label{sec:insitu:traj2}

\begin{figure*}[!t]
    \centering
    \includegraphics[width=0.96\textwidth]{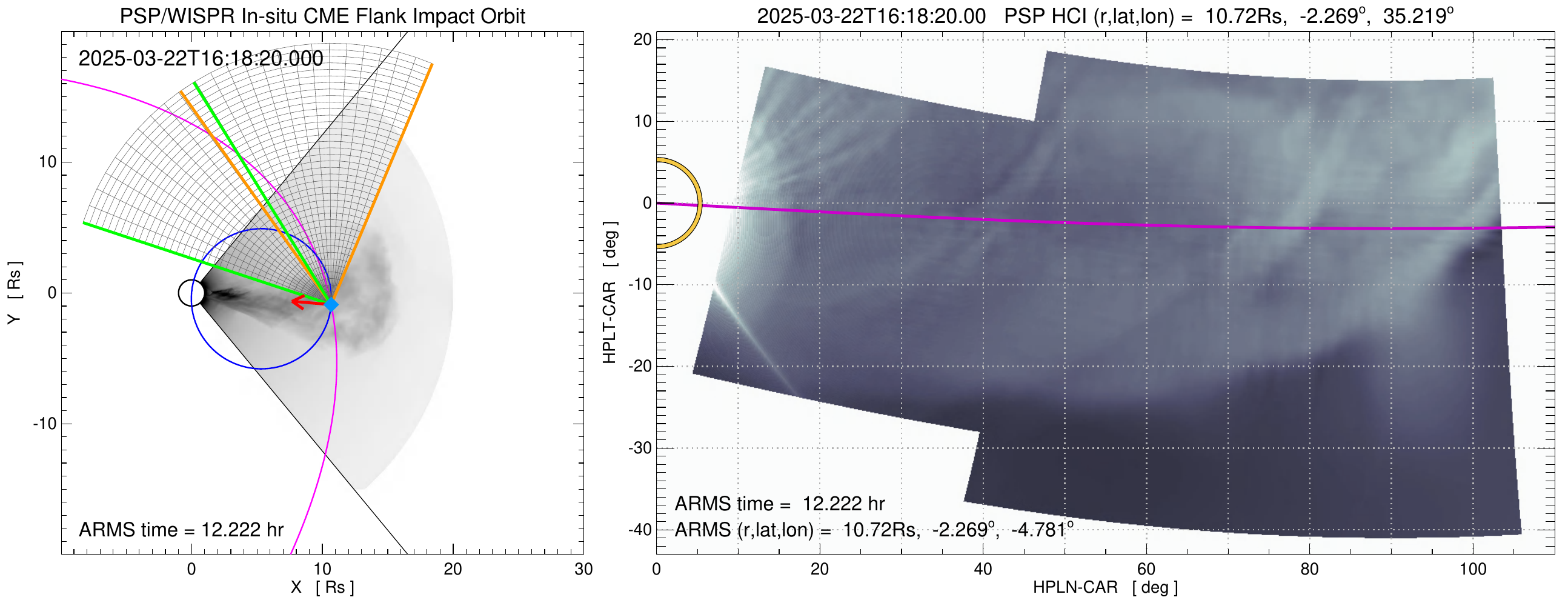}
    \includegraphics[width=0.96\textwidth]{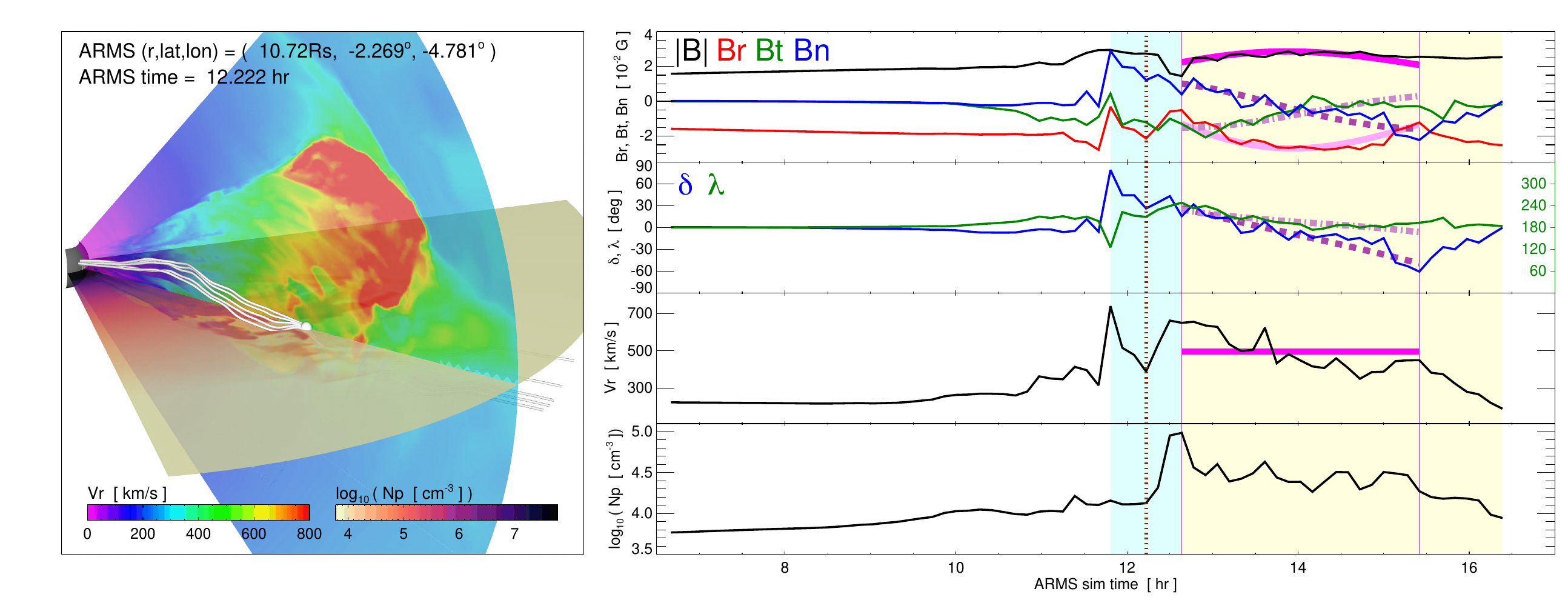}
    \caption{Synthetic PSP observations for the in-situ flank encounter with the pseudostreamer CME (Trajectory~2 orbit) in the same format as Figure~\ref{fig:psp1}. \\(An animation of this figure is available.)}
    \label{fig:psp2}
\end{figure*}

The second PSP trajectory is chosen to intersect the CME flank by lining up the ARMS simulation time $t=0$~hr with the PSP ephemeris date and time of 2025-03-22 10:45:00~UT---this is eight hours later than Trajectory~1, so our synthetic observation window is just shifted that amount further along the Encounter~23 orbit. 
Figure~\ref{fig:psp2} shows the synthetic remote-sensing and in-situ observations for the flank encounter in the same format as Figure~\ref{fig:psp1}. 
Through $t \sim 11$~hr, the evolution of the CME in the flank encounter WISPR imagery is similar to the remote-sensing trajectory, essentially confined to the upper third of the window above the ecliptic plane. As the observing spacecraft approaches the oncoming eruption for $t > 11$~hr, the CME width expands rapidly across the field of view taking on the shape of a curved, dense front with additional arc-like substructure in the white light intensity until the in-situ impact with this density front at $t=12.50$~hr as part of the CME sheath region, as seen in the lower right panel of Figure~\ref{fig:psp2}.
The CME-driven sheath region is denoted with the cyan shading (11.806~hr~$\le t <$~12.639~hr), starting at the shock and through the beginning of what looks like a moderately-structured magnetic obstacle which continues for the duration of simulation (yellow shading through $t=16.389$~hr). The sheath region begins with a magnetic discontinuity, largely in the $B_N$ component, at the exact shock arrival of the $\Delta V = 424$~km~s$^{-1}$ velocity jump.

The 3D visualization of the MHD data (Figure~\ref{fig:psp2}, lower left) shows that the CME's $V_R$ and $N_p$ signatures are much broader, and the observer's trajectory intersecting its southern flank. The CME eruption appears to be primarily aimed $\sim$15$^{\circ}$ north of the ecliptic. During the ejecta encounter, the instantaneous magnetic connectivity for our synthetic spacecraft exhibits considerable dynamic behavior, as there are visible connectivity changes between the periods of shocked and swept-up sheath region material and the extended interval of some portion(s) of the CME magnetic ejecta.

Adopting the approach of \citet{Lynch2022}, we attempt to characterize the MHD CME ejecta interval with an analytic in-situ flux rope model. 
We have used the expanding linear force-free, constant-$\alpha$ Lundquist cylinder model combined with a standard least-squares error minimization optimization for the set of fit parameters \citep[see][and references therein]{Marubashi2007,Yu2022}.
%
%
\editone{While this particular flux rope model retains the majority of the static Lundquist solution's limitations, it is able to better characterize a subset of in-situ CME events with asymmetric field magnitude enhancements via the temporal evolution introduced by self-similar expansion \citep[e.g.][]{Palmerio2024}.}
The expanding Lundquist flux rope has the usual 5 free parameters of the static cylindrical case---the 3D orientation of the symmetry axis $\left(\Theta_0, \Phi_0\right)$, the impact parameter $p_0$, the chirality $H=\pm1$, and the field magnitude on the cylinder axis $B_0$---as well as an additional timescale parameter that governs the (normalized) self-similar expansion rate, $\tau_{\rm exp}$.   

In the synthetic in-situ profiles (Figure~\ref{fig:psp2}, lower right), the vertical purple lines show the time series sub-interval used for the flux rope model fit to the large-scale field rotation. The flux rope model solutions for $|\mathbf{B}|$ (solid magneta), $B_R$ (solid pink), $B_T$ (dot-dashed magenta), and $B_N$ (dashed purple) are shown over their respective magnetic field components. The latitude and longitude of the field unit vector $\mathbf{B}/|\mathbf{B}|$ are shown as the purple dashed and magenta dot-dashed lines, respectively. The flux rope model velocity which includes the time-dependent $\mathbf{v}_{\rm exp}$ component is shown over the bulk radial velocity (note the essentially flat, non-expanding model $V_R$ solution). 
\editone{The flux rope model best-fit parameters and parameter-derived quantities are listed in Table~\ref{table:s52} and these yield a directional error metric of $\chi^2_{\rm dir} = 0.238$ for Trajectory~\#2. }
\citet{Lynch2003} used $\chi^2_{\rm dir} > 0.200$ as the quantitative threshold for a poor/unphysical flux rope fit, which is clearly the case here as our resulting flux rope size ends up being approximately zero \editone{($R_c \approx 0$~au)} due to the lack of a projected $V_R$ component across the model cylinder's circular cross-section. 

{}
{}
\begin{table}[h]
\begin{tabular}{|ll|c|c|}
     \hline
     Flux Rope Model & & Trajectory & Trajectory \\
     Parameters & & \#2 & \#3 \\
     \hline
     Elevation~angle     & $\Theta_0$ & $-0.1^{\circ}$  & $-28.2^{\circ}$ \\
     Azimuth~angle       & $\Phi_0$   & $179.9^{\circ}$ & $117.3^{\circ}$ \\
     Impact~param. & $p_0/R_c$  & $+0.50$ & $-0.88$ \\ 
     Chirality     & $H$        & $+1$ & $+1$ \\
     Mag.~field    & $B_0$      & 5.74 ${\rm \mu T}$ & 7.54 ${\rm \mu T}$ \\
     Exp.~timescale & $\tau_{\rm exp}$ & 36.4~hr & 42.3~hr \\
     \hline
     FR~radius     & $R_c$ & 0.000~au & 0.030~au \\
     Exp.~velocity & $v_{\rm exp}$ & 0.05 km~s$^{-1}$ & 22.6 km~s$^{-1}$ \\
     \hline
     Direction error & $\chi^2_{\rm dir}$ & 0.238 & 0.147 \\
     Magnitude error & $\chi^2_{\rm mag}$ & 11.91 & 37.94 \\
     \hline
\end{tabular}
\caption{The best-fit expanding Lundquist flux rope model parameters and derived quantities for the two CME-impacting synthetic trajectories. \label{table:s52}}
\end{table}

We want to emphasize that, while these parameters are not a realistic \emph{flux rope fit}, they are also not exactly a terrible interpretation given the geometric limitations of the infinite cylinder model. For example, a mostly radial field with a small apparent $B_T$ and $B_N$ rotation would likely return a $\Theta \approx 0^\circ$, $\Phi \approx 0^\circ$ or $\pm 180^\circ$ orientation which \editone{is characteristic of} a classic, high-impact parameter flank or CME-leg encounter scenario---see \citet{Lynch2022} for an expanded discussion of both these problematic in-situ encounter orientations. The relative magnitude of the rotation component will largely determine the impact parameter $p_0$. Therefore, while we cannot claim this flux rope model is a good representation of the large-scale MHD field structure, this fit is ``reasonable'' in its ``unreasonableness,'' especially when compared to the real magnetic field connectivity and its change of character between the sheath region (light cyan shading) and the CME ejecta region (light yellow shading).

\subsection{Trajectory 3: Central Encounter with CME}
\label{sec:insitu:traj3}

\begin{figure*}[t]
    \centering
    \includegraphics[width=0.96\textwidth]{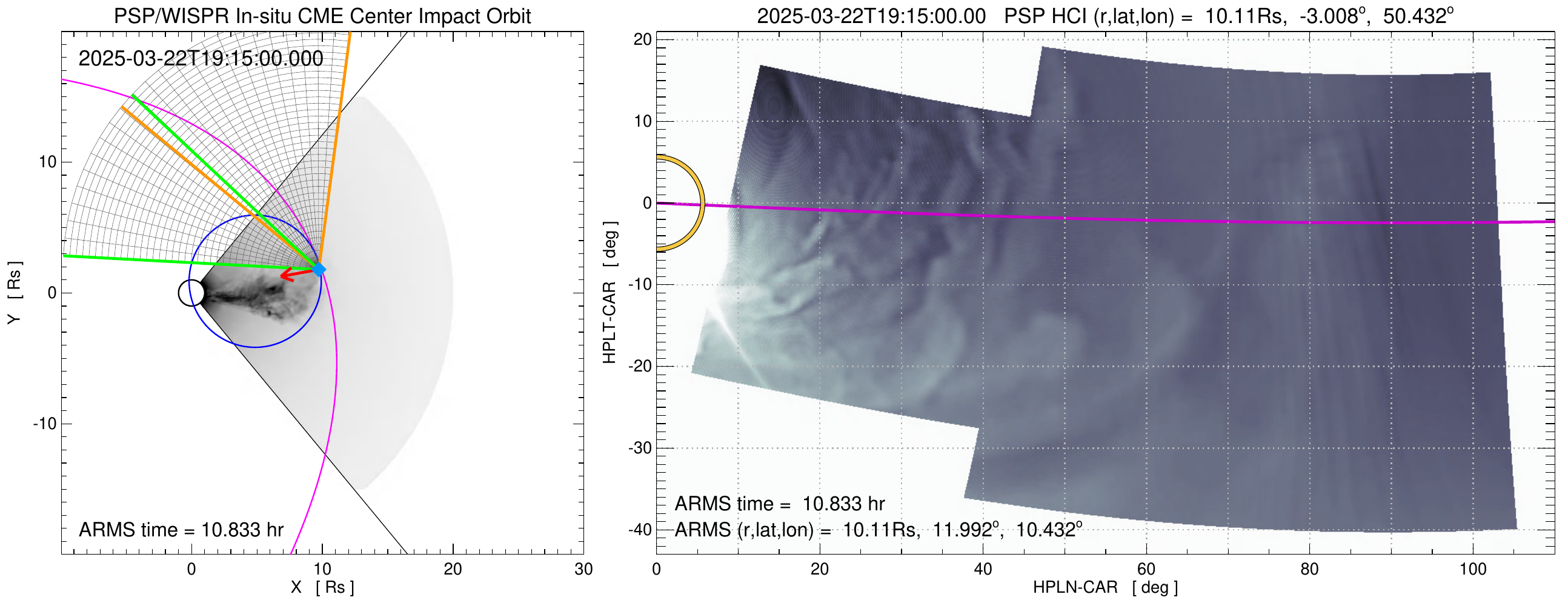}
    \includegraphics[width=0.96\textwidth]{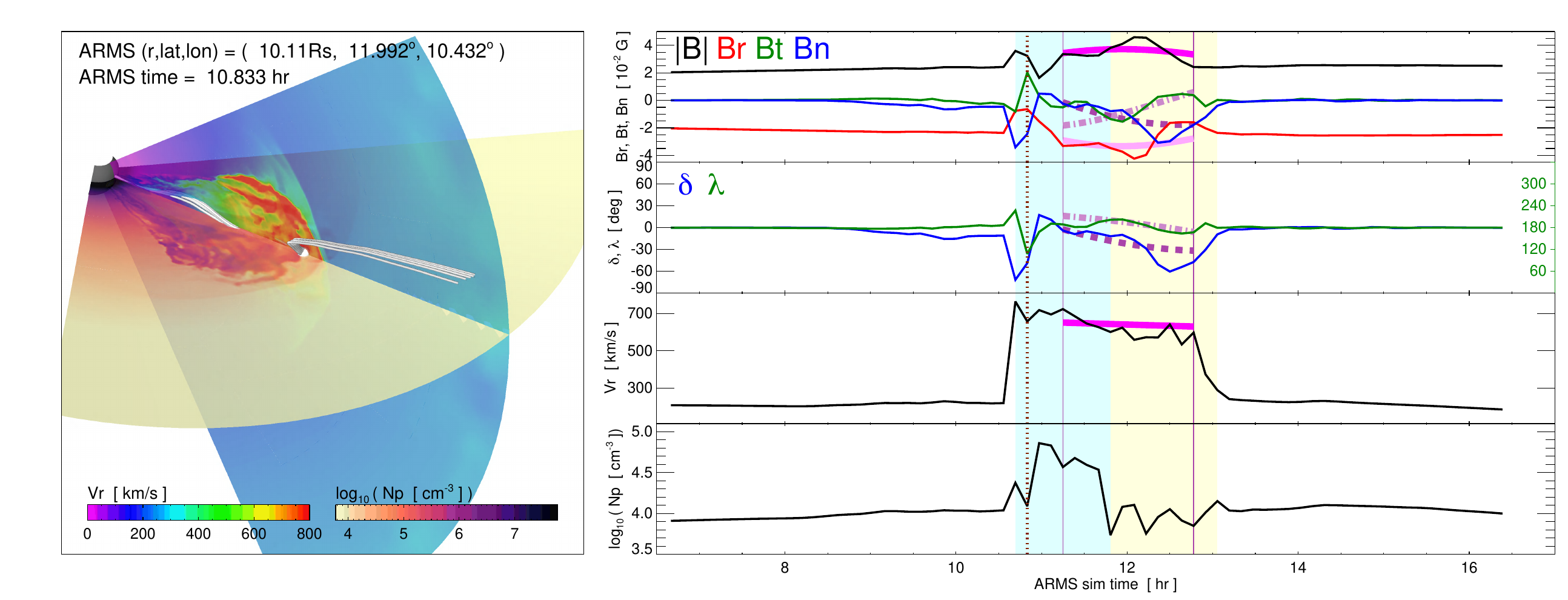}
    \caption{Synthetic PSP observations for the in-situ central encounter with the pseudostreamer CME (Trajectory~3 orbit) in the same format as Figure~\ref{fig:psp1}. \\(An animation of this figure is available.)}
    \label{fig:psp3}
\end{figure*}

The third PSP trajectory yields a central impact with the cavity/flux rope region of the CME. In addition to the temporal offset, setting ARMS $t=0$~hr with the PSP time 2025-03-22 15:05:00~UT, and the $40^{\circ}$ longitudinal offset, we apply a constant latitudinal offset of $\theta_{\rm ARMS}(t) = \theta_{\rm PSP}(t) + 15^\circ$. This has the impact of rotating the ARMS simulation with respect to the synthetic PSP orbit---now the CME enters the WISPR field of view ``from below.'' 
Figure~\ref{fig:psp3} shows the synthetic remote-sensing and in-situ observations for the central encounter in the same format as Figure~\ref{fig:psp1}. 

The central encounter trajectory has the most head-on collision between the leading edge/CME nose and the synthetic observer spacecraft. The CME becomes visible in the WISPR-Inner field of view at $t=9.583$~hr, fills the entire combined Inner and Outer view by $t=10.833$~hr---just after the shock arrival (a jump of $\Delta V \gtrsim 450$~km~s$^{-1}$ in the in-situ $V_R$ profile). The WISPR intensity enhancement/saturation fades over the next hour and by $t=11.806$~hr, the in-situ density profile drops along with the beginning of coherent rotations in the $\mathbf{B}$ components, indicating in sampling of the magnetically-dominated portion of the pseudostreamer CME ejecta. All of the synthetic WISPR imaging is of the dense sheath region of accumulated/piled-up material, which has the light cyan shading in the in-situ profiles. Immediately following the sheath is an interval with strong flux rope-like signatures in the in-situ profiles, indicated here with the light yellow shading. 

The lower left panel of the animated version of Figure~\ref{fig:psp3} shows the magnetic field line connectivity dynamism is even more dramatic than in the flank encounter. Here, because we are sampling the central strong-field cavity region, the synthetic observer passes through a portion of the CME flux rope fields that were still carrying the large-scale deflections/bends indicative of having been part of the CME leg-disconnection. The observer-connected CME field lines during the yellow-shaded interval (11.806~hr~$\le t <$~13.056~hr) are seen to wrap back around in an organized, coherent fashion, but still show the propagation and dispersion of the complex magnetic structure introduced by the leg-disconnection interchange reconnection.

Just as in Figure~\ref{fig:psp2}, we also show the flux rope model best-fit solution to the large-scale field rotations for each of the field components and the unit vector pointing.  
%
%
%
The application of the same expanding Lundquist cylindrical flux rope model yields \editone{a set of best-fit parameters, derived quantities, and error metrics that are given in Table~\ref{table:s52} for Trajectory~\#3.}
Again, this is not an especially compelling flux rope fit but the \editone{$\chi^2_{dir} = 0.147$} error metric is at least passable (i.e.\ significantly improved over the Section~\ref{sec:insitu:traj2} fit). The large-scale symmetric axis orientation \editone{of $( \Theta_0 \, , \Phi_0 ) = ( -28.2^{\circ}\,, 117.3^{\circ})$} is more reasonable {than the Trajectory~\#2 fit}, however the large value of the impact parameter \editone{$(|p_0| = 0.88R_c)$, is suggestive of some real or potential difficulties with the model solution, especially given the actual sampling position of an approximately \editone{$|p_0| \approx 0$} central encounter.}  Similar to the flank encounter time series fit, the large-scale geometry of the in-situ flux rope cylinder is obviously an incorrect starting point. Therefore, while the local field rotation may be adequately described by these analytic curves, we encourage particular caution in trying to infer anything more substantial from the CME's magnetic structure, e.g.\ its true 3D shape, twist profile, aerodynamic drag forces, quantitative magnetic flux and helicity content, etc, in cases for events fit of equal or worse quality. 
It is also interesting that despite the large velocity jump, the CME ejecta interval shows very little derived expansion in the flux rope fits, even though the profiles here are ``younger'' and exhibit more general magnetic complexity than those examined in \citet{Lynch2022}. However, we note at these distances and CME ages, the expansion dynamics can be especially difficult to disentangle from single spacecraft observations \citep{Regnault2023,Regnault2024}.

\subsection{\editone{Comparative WISPR Forward-modeling and an Application to Helical Structure in Post-CME Outflows}} \label{sec:insitu:wispr}
 
In this section, \editone{first, we summarize different impacts of the observer--CME geometry on the WL structure and evolution in the synthetic WISPR imagery. Next,} we present an interpretation of fine-scale structure in real PSP/WISPR observations made possible with the aid of our synthetic EUV and WL foward-modeling. In other words, by bringing together the simulation results from $\S$\ref{sec:interp:three} and $\S$\ref{sec:insitu:traj1}, we are able to strengthen a favorable qualitative comparison of WL structure with the inclusion of its corresponding plausible, reconnection-generated origin scenario, successfully modeled with an MHD simulation.

Line-of-sight integration effects and the time-dependent evolution of the Thomson scattering geometry for spacecraft significantly closer than 1~au often require a multi-spacecraft observational analysis for robust triangulation of the real position, propagation direction, and velocity evolution of CME transients. 
This information is critical to deproject the radial distances (and thus, velocity and acceleration profiles) of tracked features \citep[e.g.][]{Braga2021,Liewer2021,Liewer2024,Ascione2024,Cappello2024}, especially since CME trajectories are often observed to deviate from radial propagation from the source within the coronagraph field of view \citep[e.g.][]{Liewer2015,Kay2015b,Kay2016}. 
\editone{On the largest scales, the different viewing configurations of our synthetic observer orbits in $\S\S$\ref{sec:insitu:traj1}--\ref{sec:insitu:traj3} have two primary impacts on our forward-modeled WISPR WL structures and their evolution through the simulated cameras' fields of view. 
The Trajectory \#1  (Fig.~\ref{fig:psp1}) animation gives the most detailed view of the internal substructure when those density signatures are approximately at the location of the Thomson sphere. Trajectory \#2 (Fig.~\ref{fig:psp2}) animation shows the clearest transition between a remote, edge-on view of the CME to having a significant component of the transient propagation being toward the spacecraft, resulting in certain WL fronts developing a large $\theta_y$-extent with their increasing $\theta_x$ position.
There is also an asymmetry in the dynamics of the collimated leading-edge CME front arcs that develops, depending on how the observing spacecraft intersects the ejecta. For example, when the CME is largely \emph{above} our synthetic PSP observer (Trajectory \#2; Fig.~\ref{fig:psp2} animation), the leading-edge arcs become more vertical in the WISPR-O field of view with an apparent counterclockwise rotation/sweeping around an upper-right center point, whereas when the CME is \emph{below} the observer (Trajectory \#3; Fig.~\ref{fig:psp3} animation), the leading edge arcs become more horizontal in the WISPR-O field of view with apparent clockwise rotation, sweeping around a lower-left center point.}

\begin{figure}[t]
\centering
\includegraphics[width=0.96\textwidth]{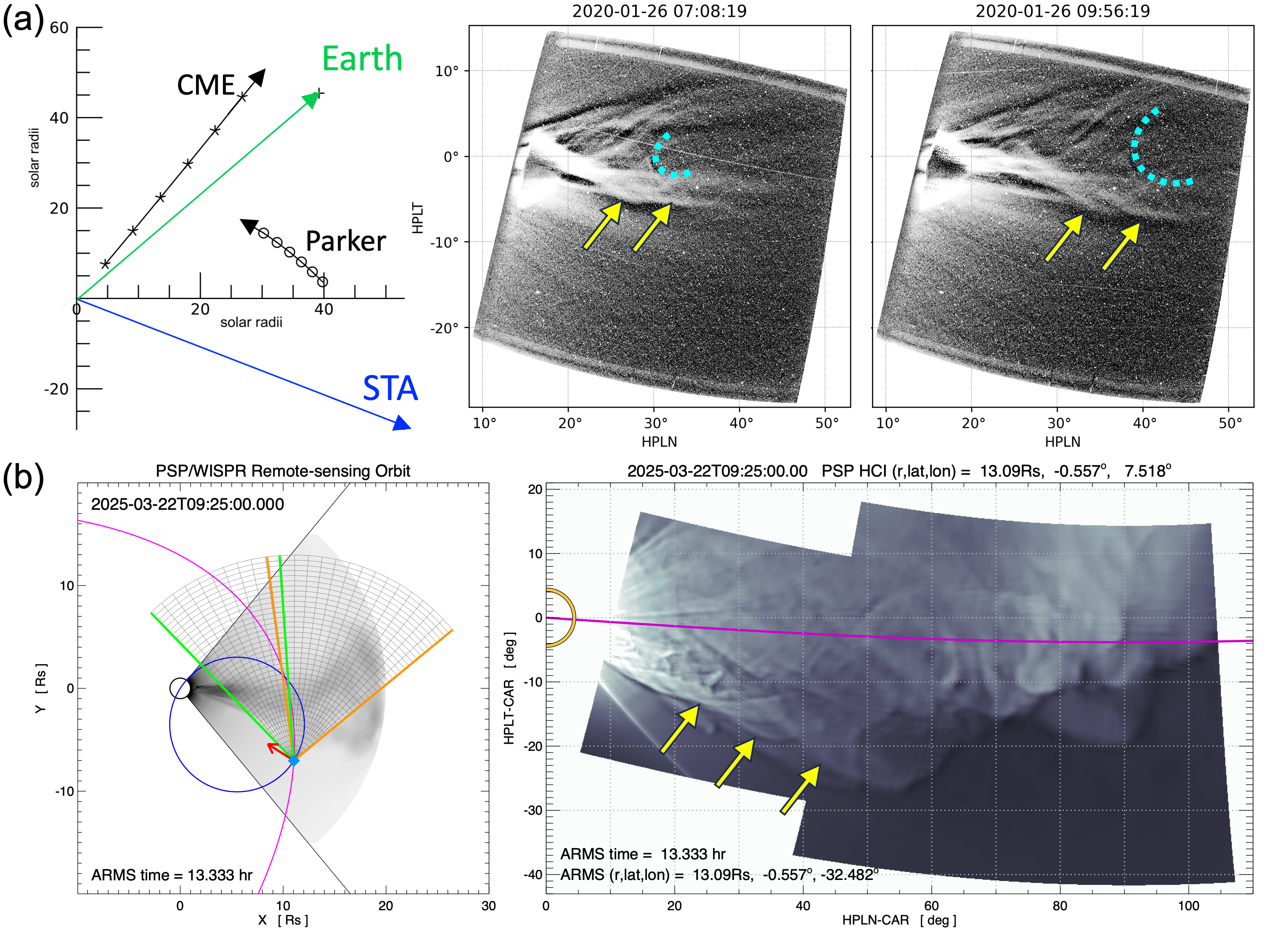}
\caption{(a) Analysis of the 2020 Jan~26 streamer blowout CME propagation direction from PSP/WISPR-I and STA (left) and two helioprojected frames of fine-scale white light structure trailing the flux rope CME \citep[adapted from][]{Liewer2021}. (b) An analogous CME-viewing geometry from the synthetic PSP/WISPR observations during the Trajectory~1 remote sensing orbit. Yellow arrows identify similar, post-CME reconnection jet-associated white light outflow structure.}
\label{fig:cf}
\end{figure}

Figure~\ref{fig:cf}(a), adpated from \citet{Liewer2021}, shows the 3D determination of the 2020 Jan~26 CME observed by PSP WISPR-I and STEREO-A (left) along with two of the PSP/WISPR-I snapshots (middle and right). 
During this event, the CME propagation direction was almost perpendicular to PSP's orbital approach, which enables maximum coverage of the event from the remote-sensing perspective. In Figure~\ref{fig:cf}(b) we show a snapshot of the synthetic PSP/WISPR observations at $t=13.333$~hr from Figure~\ref{fig:psp1}. The comparison between the overall geometry of our Trajectory~1 synthetic observer approach and PSP's viewpoint for the 2020 Jan~26 CME is favorable.
The right panel of Figure~\ref{fig:cf}(a) shows two WISPR-I images in helioprojective coordinates of the structured, post-CME outflows.
Here, the dashed cyan arc annotation indicates the C-shaped concave-up feature commonly associated with the trailing part of the flux rope cavity and used by \citet{Liewer2021} in the triangulation of the CME's 3D position. The yellow arrows point to narrow, curved white-light substructures that appear to wrap around each as part of the broader outflow enhancement. By the second frame (2020-01-26 09:56:19~UT) the post-CME outflow has almost reached the edge of the WISPR-I field of view with the wrapped, helical threads still clearly identifiable at elongation angles of $\theta_x \sim 30-40^{\circ}$. In the right panel of Figure~\ref{fig:cf}(b) we use the yellow arrows to highlight a set of consecutive curved, linear fronts at the leading edge of the post-CME outflow of our MHD simulation at nearly identical WISPR-I elongation angles. 
This outflow in our simulation's synthetic PSP/WISPR imagery corresponds to the interchange reconnection jet shown in Figure~\ref{fig:rxnfls}(c), (d) and for simulation times $t \gtrsim 10.3$~hr in the animated version of Figure~\ref{fig:wl0}.

\section{Summary and Conclusions} \label{sec:disc}

In this paper, we have extended the analysis of the \citetalias{Wyper2024} MHD simulation of a CME eruption from an idealized coronal pseduostreamer source region through the generation of synthetic remote-sensing and in-situ observations.
We have shown that MHD reconnection processes can generate an incredible amount of detailed structure and substructure in the reconnection outflows which have signatures in synthetic EUV and white light imagery.
Our analysis also demonstrates the utility of using generic, idealized simulation results to compare to specific features of observed events, despite no attempt being made to fine-tune the model parameters to better match any specific set of observations. Understanding the generic or universal features of the modeling means likely understanding (at least to zeroth-order) the main physics of the real system when there is sufficient general agreement.

The embedded magnetic dipole within a background unipolar open field is a completely generic topology and are associated with eruptive transients over decades of energy and size scales (\citealt{Sterling2016}; \citetalias{Wyper2024})---from mini-filament jetlet eruptions \citep{Sterling2021,Kumar2022,Kumar2023} through chromospheric, coronal, and X-ray jets \citep{Shibata1996,Shibata2007,Pariat2016,Raouafi2016,Wyper2017,Kumar2018,Kumar2019a}, to circular-ribbon flares \citep{Wyper2016b,Masson2019} and extended, high-latitude filaments under large-scale pseudostreamers \citep{Wyper2021,Pellegrin-Frachon2023}. 
Given that our pseudostreamer source region was fairly compact, it will be interesting to compare our results to the morphology and evolutionary dynamics of CMEs from more asymmetric or elongated pseudostreamers above extended photospheric flux distributions, such as the source region for the 2010 August~1 sympathetic eruptions from two large-scale ``parallel'' filament channels that were successfully modeled by \citet{Torok2011}. 
One of the advantages of such an idealized system is, for example, the periods of pre- and post-CME interchange reconnection in our simulation can be related to the fundamental response of an MHD system to a stressed 3D null point at the intersection of the separatrix fan and the spine field lines \citep{Syrovatskii1981,Antiochos1996,Edmondson2017,Wyper2022}.
At the same time however, even this basic topological complexity enables an energized MHD system to use magnetic reconnection to create small-scale structure to resolve the constraints imposed by the external driving \citep[e.g.][]{Wyper2014a,Wyper2014b,Lynch2016a}.

We \editone{have presented an application} of our simulation forward modeling--data analysis approach with the comparisons made in Section~\ref{sec:insitu:wispr} and Figure~\ref{fig:cf} between a set of WISPR observations to features in our idealized pseudostreamer CME eruption. 
We have shown that our fine-scale, synthetic PSP/WISPR WL structure originates in the extended, late-phase/post-CME reconnection outflow which causes the structured density enhancements along magnetic field lines tracing the remnants of the eruptive CME flux rope's large-scale twist component. 
A final observational example would be the 2022 Feb 16 CME analyzed by \citet{Palmerio2024}. Their Figure~12 (un-rotated WISPR-Inner images) show the incoming CME front with large-scale curvature, developing a succession of striation enhancements, as well as a complex internal structure, which ended up intercepting the spacecraft as the onset of a flank CME encounter. There are many generic properties of this CME encounter that are reproduced with the synthetic PSP observations for our Trajectory~2 flank encounter shown in Figure~\ref{fig:psp2} and its animation, including the incoming curved CME front and elements of its internal substructure in WL as well as the relatively weak rotations in each of the magnetic field component time series  \citep[cf.\ the in-situ time series and flux rope fits in Figure~13 of][]{Palmerio2024}. 
Lastly, we have seen how the CME flux rope disconnection and post-CME interchange reconnection jetting contribute to the transition of our PS CME from a well-structured, classic 3-part CME to a more diffuse and ``unstructured'' ejecta morphology. 
Specifically, the CME leg disconnection occurs at $r \sim 3R_\odot$ and our synthetic, limb-enhanced EUV and WL signatures showing the observational consequences of this reconnection is encouraging for current and future imaging of the extended solar corona \citep{West2023} with e.g., Solar Orbiter's EUV \citep{Rochus2020} and coronagraph \citep{Antonucci2020} imaging, novel GOES SUVI off-limb pointing configurations \citep{Seaton2021}, or upcoming missions such as PUNCH \citep{DeForest2022} and PROBA~3 \citep{Shestov2021}.


\begin{acknowledgments}

The authors would like to thank S.~K.~Antiochos, C.~R.~DeVore, and J.~T.~Karpen for their leadership in the development and use of the ARMS code, and acknowledge the 2023 and 2024 ARMS Users' Meetings as the inspiration and motivation underlying the development and write-up of these analyses. B.J.L.\ and P.F.W.\ thank the team at Predictive Science Inc.\ for hosting our visit during summer 2023 and for the fruitful discussion during the preparation of the manuscript.
B.J.L.\ acknowledges support from NSF AGS 2147399 and NASA 80NSSC24K1916, 80NSSC22K0674, 80NSSC24K1121, and 80NSSC21K1325.
P.F.W. acknowledges support from an STFC (UK) consortium grant ST/W00108X/1 and a Leverhulme Trust Research Project grant.
E.P.\ acknowledges support from NASA's PSP-GI (grant no.\ 80NSSC22K0349) and LWS (grant no.\ 80NSSC24K1108) programs as well as the NASA Parker Solar Probe Program Office for the WISPR program (under contract no.\ NNG11EK11I to NRL and subcontract no.\ N0017324C004 to PSI).
\end{acknowledgments}

%






\appendix

\section{Generation of Synthetic Observables from MHD Simulation Data} \label{sec:method}

For each of the synthetic remote-sensing observable images discussed in the following sections, we use a 3D Cartesian grid to sample the necessary MHD variables. An arbitrary position in the ARMS simulation domain $\boldsymbol{r}$ can be written as $\boldsymbol{r}= \boldsymbol{r_0} + \ell \, \boldsymbol{\hat{e}_3}$ where we define $\boldsymbol{r_0}$ as the position in the plane of the sky (the synthetic image plane; $\boldsymbol{r_0} = x' \, \boldsymbol{\hat{e}_1} + y' \, \boldsymbol{\hat{e}_2}$) and the coordinate $\ell$ along the line of sight (LOS; $\boldsymbol{\hat{e}_3}$) perpendicular to the image plane. The transformation from the native ARMS spherical coordinates $(\boldsymbol{\hat{r}},\, \boldsymbol{\hat{\theta}},\, \boldsymbol{\hat{\phi}})$ into the viewing-perspective Cartesian coordinates $(\boldsymbol{\hat{e}_1},\, \boldsymbol{\hat{e}_2},\, \boldsymbol{\hat{e}_3})$ is achieved via the standard spherical-to-Cartesian transformation $(\boldsymbol{\hat{x}},\, \boldsymbol{\hat{y}},\, \boldsymbol{\hat{z}})$ followed by the application of the 3D Euler angle rotation matrix $\mathbb{R}$.
In practice, we specify the synthetic observable Cartesian grid coordinates, apply the inverse rotation $\mathbb{R}^{-1}$, and then convert the resulting Cartesian coordinates back into their native spherical coordinates to sample the ARMS variables and/or perform the relevant calculations based on the ARMS variables.
In Figures~\ref{fig:euv} and \ref{fig:wl0}, we derive the synthetic remote-sensing observable signatures from three primary viewpoints: \textit{Pole}---viewing perspective from the northern solar pole toward $-\boldsymbol{\hat{z}}$ (plane of the sky in $\boldsymbol{\hat{x}}$, $\boldsymbol{\hat{y}}$); \textit{Limb}---viewing perspective from the equator toward $+\boldsymbol{\hat{y}}$ (plane of the sky in $\boldsymbol{\hat{x}}$, $\boldsymbol{\hat{z}}$); and \textit{Disk}---viewing perspective from the equator toward $-\boldsymbol{\hat{x}}$ (plane of the sky $\boldsymbol{\hat{y}}$, $\boldsymbol{\hat{z}}$).

\subsection{Extreme Ultraviolet Emission} \label{euvmath}

Quantitative calculation of synthetic EUV emission intensities requires a moderately sophisticated MHD energy equation including field-aligned thermal conduction, radiative losses, and a source term from a coronal heating model. As the synthetic EUV emission reflects both density and temperature structure, it can be a valuable tool for assessing the performance and robustness of coronal heating models \citep{Lionello2009, Downs2010, vanderHolst2014, Mikic2018} or energy transfer during CMEs and other eruptive transients \citep[e.g.][]{Jin2017b,Szente2017,Reeves2019}.
For a more detailed description of the elements that go into a comprehensive, thermodynamic MHD model and forward modeling synthetic emission, we refer the reader to the excellent presentation in \citet{Downs2010}.

In general, the synthetic EUV intensity at the plane-of-the-sky position $\boldsymbol{r_0}$ is calculated by integrating along the LOS direction $\boldsymbol{\hat{e}_3}$,
\begin{equation}
    I_{\rm EUV}(\boldsymbol{r_0}) = \int_{\rm LOS} d\ell \; n_e^2(\boldsymbol{r}) \; F\left( n_e(\boldsymbol{r}), T(\boldsymbol{r}) \right) \; . 
\end{equation}
Here, $F(n_e,T)$ is the instrument response function for the emission over the filter of interest's wavelength range. For a given emitting plasma parcel's contribution to the line-of-sight integral, its density and temperature determine the incoming spectrum \citep[requiring an emission model, e.g.\ typically calculated with the CHIANTI atomic spectra database;][]{Dere2023} which is then convolved with the individual instrument's filter response. The analysis routines for the majority of solar data (e.g., from missions such as SOHO, Hinode, STEREO, SDO, IRIS) are publicly available as various packages for SolarSoft IDL and increasingly Python.

However, given our MHD simulation's extremely simple energy equation (an isothermal plasma), a detailed accounting of the atomic physics of the emission line spectra and instrumental calibration factors are unnecessary as we do not need to reproduce actual detector data number counts. Thus, we calculate our synthetic EUV emission as $I_{\rm EUV} \sim \int d\ell \, n_e^2(\boldsymbol{r})$ (units of cm$^{-5}$) via the usual charge-neutrality relation with the MHD proton density $n_e(\boldsymbol{r}) = \rho(\boldsymbol{r})/m_p$. We note this EUV emission approximation is a fairly common zeroth-order estimate \citep{Robbrecht2009} but is still able to reveal complex and dynamic multi-scale structure \citep{Lynch2019,Lynch2021}. 
For each pixel $(i,j)$ in our synthetic image, the line-of-sight integral is performed by interpolating the MHD solution at positions $\boldsymbol{r}(i,j,k) = \boldsymbol{r}_0( i , j ) + k \Delta \ell \, \boldsymbol{\hat{e}_3}$  over $k = \left( 0, N_k-1 \right)$ steps along the LOS and summing the sampled electron density contributions, $n^2_e(\boldsymbol{r}) = \left( \rho(\boldsymbol{r}) / m_p \right)^2$ derived from our single-fluid proton density multiplied by the LOS step size $\Delta \ell$.
The synthetic EUV images in Figure~\ref{fig:euv}(a),(b) were constructed as 2D arrays ($1024 \times 512$) of lines of sight over the $4 R_\odot \times 2 R_\odot$ domain with 512 points along each line of sight ($\ell \in [-4R_\odot,4R_\odot]$) corresponding to a $\Delta \ell = 0.0156\,R_\odot$. The EUV image in Figure~\ref{fig:euv}(d) was constructed as a $512 \times 512$ array with 256 points along each line of sight ($\ell \in [0, 2.5R_\odot]$) corresponding to a $\Delta \ell = 0.0098\,R_\odot$.

\subsection{Thomson-scattered White Light} \label{wlmath}

Similarly, the procedure for calculating the synthetic Thomson-scattered WL intensity consists of a line-of-sight integral of the electron density weighted by a function of its position along the observer's line of sight. 
This relatively straightforward dependence on the MHD density has meant synthetic WL forward modeling of coronal streamers and CME dynamics continue to be widely used in the analysis of simulation results \citep{Vourlidas2013,Lynch2016,Lynch2021,Parenti2021,Wyper2021,Ben-Nun2023}.

The Thomson scattering efficiency is largely determined by the source--scattering point--observer geometry \citep{Billings1966}. The total WL intensity at the position $\boldsymbol{r_0}$ is then  
\begin{equation}
    I_{\rm WL}(\boldsymbol{r_0}) = \int_{\rm LOS} d\ell \; n_e(\boldsymbol{r}) \; F(\boldsymbol{r}) \; ,
\end{equation}
where we use the 2D image plane and perpendicular line-of-sight coordinates described above, and  $F(\boldsymbol{r})$ represents the Thomson scattering cross-section and geometric terms based on {\tt eltheory.pro} in SolarSoft IDL \citep[e.g. see][]{Billings1966,Hayes2001}. Using the \citet{Billings1966} definitions, 
\begin{equation}
F(\boldsymbol{r}) = \sigma_e \frac{\pi}{2}\left( 1 - \frac{u}{3} \right)^{-1} \Bigl[ \, 2\bigl[ \left(1-u\right)\mathcal{C} + u\mathcal{D} \bigr] - \sin^2{\chi}\bigl[ \left(1-u\right)\mathcal{A} + u\mathcal{B} \bigr] \, \Bigr] \; ,
\end{equation}
where the Thomson cross-section is $\sigma_e = 7.95\times10^{-26}$~per steradian, $u=0.63$ is the standard limb-darkening coefficient, and the resulting units are in mean solar brightness. The terms $\mathcal{A}$, $\mathcal{B}$, $\mathcal{C}$, and $\mathcal{D}$ are all functions of position (usually expressed in terms of the angles $\sin{\Omega} = 1/r$, $\cos{\Theta} = r_0/r$), and $\chi$ is the angle between the electron scattering point at $\boldsymbol{r}$ and the observer's LOS along $\boldsymbol{\hat{e}_3}$.

The synthetic WL image is constructed via summation over the MHD simulation's $n_e(\boldsymbol{r})$ values multiplied by its geometric scattering $F(\boldsymbol{r})$. The units of the constants are such that after the line-of-sight integration our intensity units are in Mean Solar Brightness (MSB). 
The three sets of WL images in Figure~\ref{fig:wl0} were each generated at a resolution of $512 \times 512$ with 512 points along the line of sight but over varying spatial domains: for the inner-most $3 R_\odot \times 3 R_\odot$ domain, $\ell \in [-1.5R_\odot, 1.5R_\odot]$ yielding a line-of-sight step-size $\Delta \ell = 0.0059 R_\odot$;  the intermediate $8 R_\odot \times 8 R_\odot$ domain has $\ell \in [-4R_\odot, 4R_\odot]$ and a $\Delta \ell = 0.0156 R_\odot$; and for the outermost $18 R_\odot \times 18 R_\odot$ domain, $\ell \in [-9R_\odot, 9R_\odot]$ and $\Delta \ell = 0.0352 R_\odot$. 
%

\subsection{Image Processing for Synthetic EUV and White-light Observations}
\label{imgproc}

As the densities and temperatures in the MHD models become more realistic, the forward-modeled synthetic EUV and WL observations begin to require additional image processing or enhancement procedures in much the same way as the analogous real observations. For example, using logarithmic intensity scales and/or applying radial or ``flat-fielding'' corrections are all techniques employed to pull out the features of interest. 

\begin{figure*}[!t]
    \centering
    \includegraphics[width=0.98\textwidth]{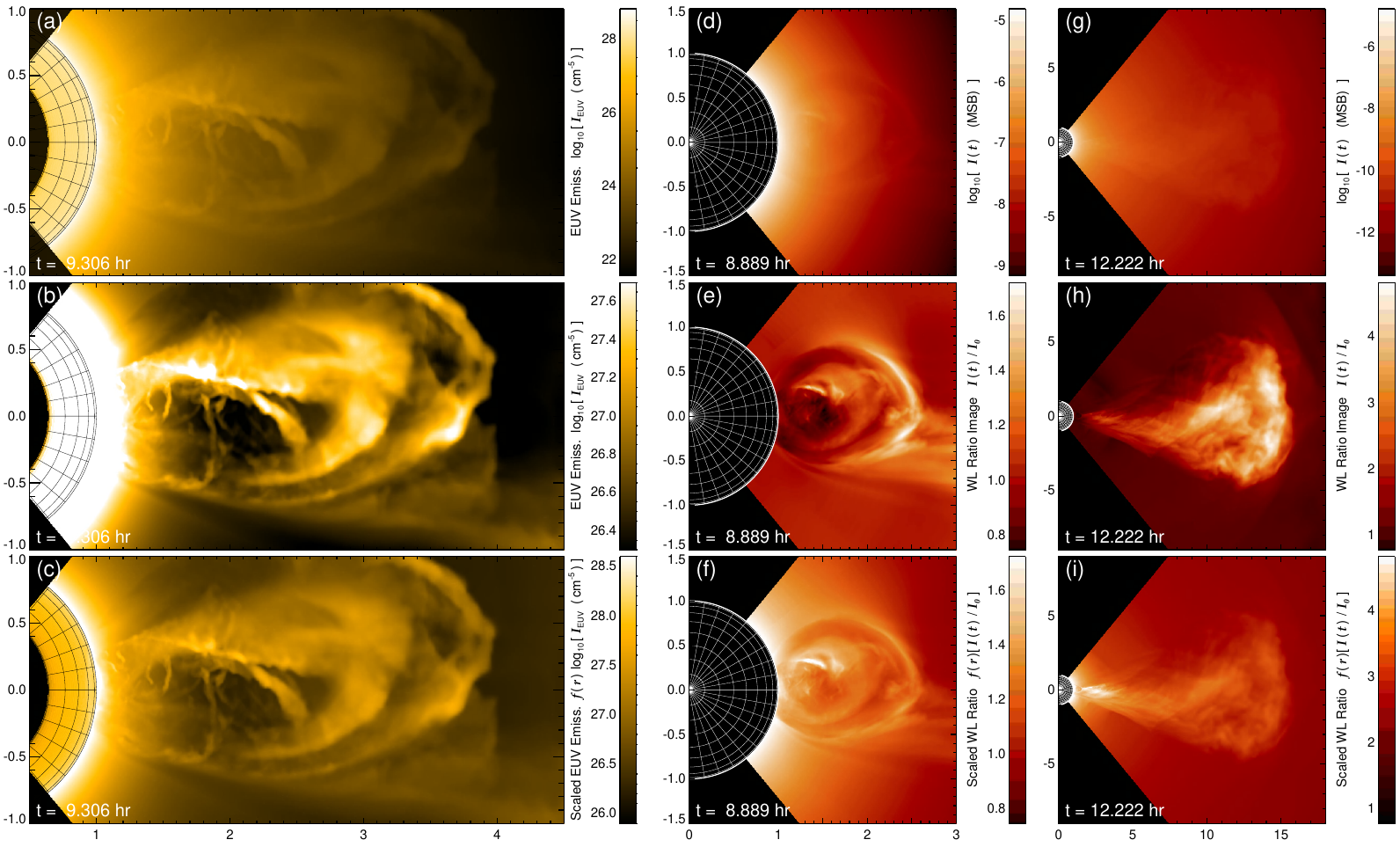}
    \caption{Imaging processing for the synthetic EUV and WL coronagraph data. Left column, \emph{Pole} view of Figure~\ref{fig:euv} showing the (a) original image, (b) an over-saturated range, and (c) with the applied radial scaling. Center and right columns, \emph{Pole} view  inner and outer coronal domains of Figure~\ref{fig:wl0} synthetic WL imaging. (d), (g) the original WL images. (e), (h) the WL ratio images $\left( I/I_0 \right)$. (f), (i), the final radially-scaled WL images.  
    \\(An animation of this figure is available.)}
    \label{fig:image2}
\end{figure*}

The top row of Figure~\ref{fig:image2} shows the \emph{Pole} view of the EUV and the inner and outer WL images displayed at (a) $t = 9.306$~hr, (d) 8.889~hr, and (g) 12.222~hr, respectively, applying a simple logarithmic scaling of the synthetic intensity values. While there is some significant off-limb structure visible in the EUV panel, the CME signal in the WL frames is extremely faint. 
The middle row of Figure~\ref{fig:image2} shows the application of a basic normalization technique in each case: (b) an aggressively clipped (over-saturated) range of the original log EUV intensity; and (e), (h) WL ratio images $I(t)/I_0$. Here $I_0$ is the WL intensity calculated from the initial, spherically-symmetric equilibrium density profile $N_p(r) = n_0 \exp{\left[-r/H\right]}$ where $H = 2 k_{B} T_0 / ( m_p g_{\odot} )$ is the scale height associated with a $T_0=1$~MK isothermal plasma with $p_0 = 0.33$~dyn~cm$^{-2}$ and $n_0 = 1.195 \times 10^9$~cm$^{-3}$. For the lower corona in panel (e), the WL ratio values typically range from 0.8--1.5 whereas for transients in the extended corona the maximum values can reach `a few,' e.g.\ $\lesssim 5$ in panel (h).
The bottom row of Figure~\ref{fig:image2} shows scaled versions of the synthetic images where we have introduced a radial function multiplier in an attempt to regularize the signal across the field of view. For EUV in panel (c), we choose the function $f(r) = (r/R_\odot)^{0.13}$ to apply to the logarithmic intensity of $I_{EUV}$ in \ref{fig:image2}(a). For WL imagery in panels (f) and (i), we choose the function $f(r) = ( 3.5\,R_\odot / r )$ to apply to the ratio images $I(t)/I_0$ in (e) and (h), respectively. We note this form of the WL $f(r)$ re-imposes a softer $r^{-1}$ intensity fall off to the ``over-flattened'' ratio images. The animated version of Figure~\ref{fig:image2} shows processing impact over the stages of the CME's evolution through each of the respective fields of view.

\section{Synthetic Parker Solar Probe Observations} \label{sec:psp}

\subsection{Parker Solar Probe WISPR Imagery} \label{sec:psp:two}

The procedures we have implemented for the generation of synthetic PSP/WISPR imagery from the ARMS simulation data essentially follow those employed by \citet{Poirier2020,Poirier2023} in their forward-modeling and analysis of MHD solar wind streamer structures. In the synthetic WISPR images presented in this paper, we used the following field of view and image resolution. The image arrays were constructed as $1100 \times 640$ lines of sight over the helioprojective coordinate range $\theta_x \in [ 0^{\circ}, 110.0^{\circ}]$ and $\theta_y \in [ -43^{\circ}, 21^{\circ}]$ with 1024 points along each LOS ($\ell \in [0.01R_\odot, 20R_\odot]$, $\Delta \ell = 0.0195R_\odot$). We note these angular ranges were chosen because they cover the maximum extent of the (predicted) WISPR-I and WISPR-O fields of view over each of the orbit intervals.

The synthetic wide-angle WL WISPR images are constructed with minimal modifications to the same procedure as employed for the static coronagraph images (Section~\ref{wlmath}). In practice, we are still constructing a 2D image plane and integrating in the third dimension, we are just using spacecraft-centered spherical coordinates $(r,\frac{\pi}{2} - \theta,\phi) = (\ell, \theta_y, \theta_x)$ rather than the Cartesian decomposition above. In this configuration, the line-of-sight integration directions are radial cuts from the PSP position, so $\boldsymbol{\hat{e}_3} = \left( \boldsymbol{r} - \boldsymbol{r}_{\rm PSP} \right) /  | \boldsymbol{r} - \boldsymbol{r}_{\rm PSP} |$, which now has a $\boldsymbol{r}_{\rm PSP}(t)$ time dependence.

The spacecraft and WISPR camera pointing are obtained from orbit date and position meta-data associated with the World Coordinate System \citep[WCS;][]{Thompson2006,Thompson2010} object created for the synthetic spacecraft position. In this way, we are creating artificial WISPR FITS headers using the same coordinate system specifications as the observational data (e.g.\ \citealt{Hess2021}; see also the WISPR User Guide\footnote{\url{https://wispr.nrl.navy.mil/sites/wispr.nrl.navy.mil/files/wispr_data_user_guide_v3.pdf}}). The geometric component of the WISPR field of view projected area is calculated on the basis of the WCS environment for the PSP date (position) along with the WISPR-Inner, WISPR-Outer instrument specifications.

Optical distortion for the detector mapping to the observational line of sight positions into helioprojective coordinates can be applied via the {\tt PV2} coefficients and the zenithal plane coordinate type {\tt ctype = [`HPLN-ZPN', `HPLT-ZPN']} designation during the WCS generation, whereas maintaining the helioprojective plate carr\'{e}e coordinate type, {\tt ctype = [`HPLN-CAR', `HPLT-CAR']}, returns the celestial sphere projection of the undistorted regular/rectangular angular arrays of the WISPR-I and WISPR-O fields of view.  For each synthetic image, the WISPR-I, WISPR-O detector boundaries are obtained from the {\tt lat(i,j)} and {\tt lon(i,j)} variables in the {\tt data1} and {\tt data2} structures of the combined image's meta-data returned by the {\tt wispr\_join\_images()} routine with the {\tt sys=`SPP\_HPC'} designation for desired coordinate system of the combined image. Each camera's field of view (FOV) mapping to the HPC coordinates as a function of the image axis position is provided in these 2D arrays. For example, {\tt data1.lon(i,j)} and {\tt data1.lat(i,j)} are the WISPR-Inner FOV position elongation and elevation angles ($\theta_x$, $\theta_y$) corresponding to the line-of-sight direction that is mapped to the pixel position $(i,j)$. Therefore, the first and last column vectors ($i=0$ and $N_x-1$) represent the left and right edges while the first and last row vectors ($j=0$ and $N_y-1$) correspond to the top and bottom edges---each of these line segements, $\mathcal{L} \left( \theta_x, \theta_y \right)$, are smooth functions of the HPC angular coordinates and everything has a time-dependence due to the rapidly changing position of the spacecraft. As illustrated in Figure~\ref{fig:image3}(a), these FOV boundaries define the ``visible area'' of the actual WISPR-I (green outline) and WISPR-O cameras (orange outline) within the larger $110^\circ \times 64^\circ$ angular range given above, and thus also define the subset of the 2D array of lines that fall outside of the WISPR FOVs and should be masked.

Figures~\ref{fig:psp1}--\ref{fig:psp3} and \ref{fig:image3} all show examples of the synthetic WISPR white-light projections with the FOV masks imposed without the additional optical distortion (see Palmerio et al.\ 2024, in~preparation, for an example of synthetic WIPSR imaging with the optical distortion applied). In each of the Figure~\ref{fig:psp1}--\ref{fig:psp3} and \ref{fig:image3} animations, the spatiotemporal evolution of the synthetic WISPR FOVs is apparent.  We note the {\tt wispr\_join\_images()} routine is included with the JPL contribution to the WISPR processing codes distributed within SolarSoft IDL.

\subsection{Simple Feature Enhancement for Synthetic PSP/WISPR Observations}
\label{enhance}

\begin{figure*}[t]
    \centering
    \includegraphics[width=0.98\textwidth]{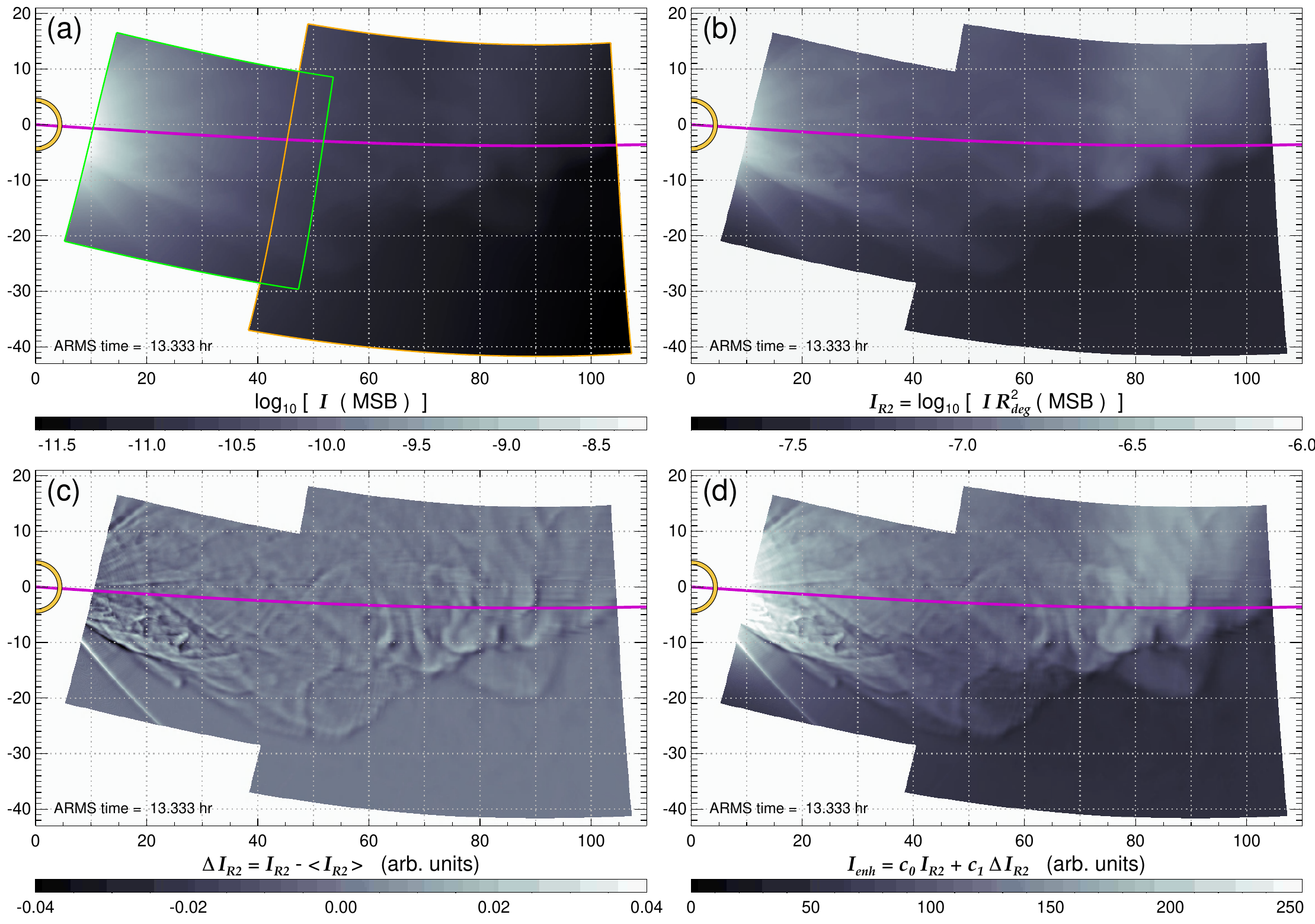}
    \caption{Synthetic PSP/WISPR image processing applied to the Trajectory 1 data. (a) Original white light intensity image $I(\theta_x,\theta_y)$ in logarithmic units of Mean Solar Brightness (MSB) with the individual Inner (WISPR-I, green outline) and Outer (WISPR-O, orange outline) camera fields of view also shown. (b) Intensity image after the application of a radial scaling factor $I R_{\rm deg}^2$ (see \S\ref{enhance}). (c) The high-pass filtered ``difference image,'' $\Delta I_{R2}$. (d) Final enhanced image as the weighted sum $I_{\rm enh} = c_0 I_{R2} + c_1 \Delta I_{R2}$. The Sun is shown as the yellow circle at (0,0) and the magenta line shows the PSP orbit plane projected into the synthetic FOV. \\(An animation of this figure is available.)}
    \label{fig:image3}
\end{figure*}

Here we describe a fairly common edge-enhancement procedure, ``unsharp masking,'' which accentuates intensity gradients and reveals fine-scale structure and substructure. We define a $\Delta I$ image that emphasizes the local variation of image values, i.e. $\Delta I = I - \langle I \rangle $ where the mean image, $\langle I \rangle$, represents a smoothed version of $I$ retaining only the lower spatial frequency content. Therefore, $\Delta I$ contains the higher spatial frequency components of the original image. The final ``edge-enhanced'' image is then constructed as the linear combination $I_{enh} = c_0 \, I + c_1 \, \Delta I$. Here we use a $22 \times 11$ pixel boxcar average as implemented in the IDL {\tt smooth()} function to create the low-frequency $\langle I \rangle$ background and values of $(c_0, \, c_1) = (0.25, \, 4.0)$. Figure~\ref{fig:image3} shows this procedure applied to the $t=13.333$~hr frame of the synthetic WISPR images from the remote-sensing encounter orbit described in Section~\ref{sec:insitu:traj1} (PSP Trajectory \#1). Figure~\ref{fig:image3}(a) shows the original composite image over WISPR-I and WISPR-O fields of view in units of MSB. Figure~\ref{fig:image3}(b) shows $I_{R2}$, the original image multiplied by the function $R_{deg}^2 = \theta_x^2 + \theta_y^2$, which is functionally an $r^2$-scaling applied to the whole image (to decrease the dynamic range of the synthetic intensity) but with the helioprojective longitude ($\theta_x$) and latitude ($\theta_y$) values given in degrees rather than a normalized radial distance. Figure~\ref{fig:image3}(c) shows the high-frequency component (fluctuation around the mean), $\Delta I_{R2} = I_{R2} - \langle I_{R2} \rangle$. Figure~\ref{fig:image3}(d) shows the edge-enhanced image $I_{enh}$ calculated from the panel \ref{fig:image3}(b) and \ref{fig:image3}(c) images. An animated version of Figure~\ref{fig:image3} is included in the online version of the article. There is a moderately straightforward extension of our \emph{ad hoc} ``unsharp masking'' prescription to the rigorous wavelet-scale decomposition image processing described in \citet{Stenborg2008}, which itself has been further modified for the `LW' processed WISPR data product that accentuates fine-scale features \citep{HowardR2022}.

\subsection{In-situ Observations from Parker Solar Probe-like Trajectories} \label{sec:psp:one}

The synthetic in-situ observations are obtained from the MHD simulation data via the same method as described in \citet{Lynch2022}. 
We used the SPICE kernels \citep{Acton1996,Acton2018} to generate the PSP Encounter 23 ephemeris data via routines in the PSP analysis packages for SolarSoft IDL. Once we have determined the orbit interval we wish to apply to our simulation data, we calculate the required temporal offset/correspondence between a reference simulation time (e.g., $t=0$) and a PSP orbit time that yields the desired trajectory interval through the domain. 
Specifically, after creating an array of the desired PSP position times to match the MHD simulation output file cadence (here $\Delta t = 500$~s), we call {\tt get\_sunspice\_lonlat()} with the designation {\tt SYSTEM = `HCI'} for heliocentric inertial coordinates (other systems are available, e.g.\ Carrington coordinates, etc). We then loop through the simulation output files to obtain an estimate of the MHD variables of interest at each of the synthetic observers' orbital positions using the standard volume-weight (3D linear) interpolation between the surrounding eight grid points. 
The resulting synthetic in-situ time series of the MHD quantities $\boldsymbol{B}_{\rm RTN}$, $V_R$, $N_p$ for each of the three observing trajectories are shown in Figures~\ref{fig:psp1}--\ref{fig:psp3} and discussed in Section~\ref{sec:insitu}.


\bibliography{apj-jour,master_psobs_0712,biblio_pw}{}
\bibliographystyle{aasjournal}



\end{document}